\documentclass[final,1p,times,authoryear]{elsarticle}

\usepackage{amssymb}
\usepackage{amsmath}
\usepackage{subcaption}
\usepackage{array}
\usepackage{enumitem}

\newtheorem{theorem}{Theorem}

\newtheorem{prop}{Proposition}

\newtheorem{cor}{Corollary}

\journal{Journal of Statistical Planning and Inference}

\begin{document}

\begin{frontmatter}



\title{Moment-based Random-effects Meta-analysis \\ Equipped with Huber's M-Estimation} 


\author[label1]{Keisuke Hanada} 
\ead{hanada.keisuke.es@osaka-u.ac.jp}
\author[label1]{Tomoyuki Sugimoto}

\affiliation[label1]{organization={Graduate School of Engineering Science, Osaka University},
            addressline={1-3}, 
            city={Machikaneyama-cho, Toyonaka},
            postcode={560-8531}, 
            state={Osaka},
            country={Japan}}

\begin{abstract}
Meta-analyses are commonly used to provide solid evidence across numerous studies. Traditional moment methods, such as the DerSimonian-Laird method, remain popular in spite of the availability of more accurate alternatives. While moment estimators are simple and intuitive, they are known to underestimate the variance of the overall treatment effect, particularly when the number of studies is small. This underestimation can lead to excessively narrow confidence intervals that do not meet the nominal confidence level, potentially resulting in misleading conclusions.
In this study, we improve traditional moment-based meta-analysis methods by incorporating Huber's M-estimation to more accurately capture the distributional characteristics of between-study variance. Our approach enables conservative parameter estimation, even when almost all existing methods lead to underestimation of between-study variance under a small number of studies. Additionally, by deriving the simultaneous distribution of overall treatment effect and between-study variance, we propose facilitating a visual exploration of the relationship between these two quantities.
Our method provides more reliable estimators for the overall treatment effect and between-study variance, particularly in situations with few studies. Using simulations and real data analysis, we demonstrate that our approach always yields more conservative results compared to traditional moment methods, and ensures more accurate confidence intervals in meta-analyses.
\end{abstract}


\begin{highlights}
\item The proposed M-estimator approach enables conservative confidence interval estimation for treatment effects, even with few studies.
\item The simultaneous density function facilitates visual assessment of overall treatment effect and between-study heterogeneity.
\item The exact distribution is derived for an estimated between-study variance under generalized moment estimation.
\item The M-estimator approach completely avoids computational failures encountered in conventional methods.
\end{highlights}


\end{frontmatter}



\section{Introduction}
\label{sec1}
Meta-analyses are essential for synthesizing evidence across multiple studies addressing similar research questions, thereby providing more solid conclusions \citep{dersimonian1986meta}. Random-effects meta-analyses introduce the between-study random variable to account for the heterogeneity unobserved among studies. This heterogeneity can stem from differences in study design, outcome measurement methods, protocol variations, and unobserved confounders. While random-effects models offer a more flexible framework compared to fixed-effect models, a more accurate estimation of the between-study variance is required to make inferences on the random-effects model more valid.

Traditional moment methods, such as the DerSimonian-Laird (DL) method, are commonly used in meta-analysis, but have been reported to have some statistical problems. For example, it has been shown that the variance of the overall treatment effect in random-effects meta-analysis is often prone to underestimation due to its expected variance falling below the Cramér-Rao lower bounds \citep{li1994bias, viechtbauer2005bias}. This underestimation of between-study variance is especially common when the number of studies is small, resulting in excessively narrow confidence intervals (CIs) that do not achieve the nominal confidence level. As a consequence, overconfident and potentially erroneous conclusions may be drawn, such as incorrectly deeming ineffective treatments effective, which can have serious implications for clinical decision-making. For these reasons, to mitigate the underestimation of between-study variance in cases with a small number of studies, more conservative estimation methods have been developed \citep{hartung1999alternative, rover2015hartung, michael_exact_2019}. Among these methods, the methods of \cite{michael_exact_2019} and \cite{hanada2023inference} offer theoretically exact interval estimation, they often require tuning parameters to achieve accuracy and may still yield overly conservative estimates in practical applications. The Hartung-Knapp-Sidik-Jonkman (HKSJ) approach can be flexibly applied regardless of the specific method used to estimate the between-study variance \citep{hartung1999alternative}, but may suffer from lower coverage probabilities than the nominal level when studies with larger or smaller sample sizes are included. The modified Knapp-Hartung method \citep{rover2015hartung} introduces corrections to improve the coverage, thereby producing more reliable and conservative CIs, even when the HKSJ approach may result in under-estimate.

While the methods of \cite{michael_exact_2019}, HKSJ, and \cite{rover2015hartung} are appropriate for conservative interval estimation of the overall treatment effect, they cannot estimate the variability of the between-study variance. Meta-analysts need to select the estimation method for between-study variance independently of the estimation of the overall treatment effect. Although variance estimation for between-study variance using generalized moment estimators has been proposed \citep{sidik2022quantifying}, these methods address the estimation of between-study variance and overall treatment effect separately, without considering their interrelationship. Therefore, this issue could be addressed by developing a meta-analysis method that simultaneously accounts for the overall treatment effect and between-study variance.

In this study, we extend traditional moment estimation methods by incorporating Huber's M-estimation framework \citep{huber1992robust}. This approach (i) improves the coverage probability for the overall treatment effect and brings it closer to the nominal level compared to traditional methods, and (ii) provides the distribution of between-study variance for each moment estimator. In this paper, we focus on the discussion for constructing the CI, leaving to \cite{hanada2023inference} who discuss that their almost-exact distributional approach can test the overall treatment effect at the nominal level. By applying the M-estimation, we account for the uncertainty in between-study variance, which is often overlooked in traditional methods. This framework allows for the simultaneous estimation of the overall treatment effect and between-study variance, providing insights into the relationship between these two quantities. Especially, by deriving the simultaneous distribution of overall treatment effect and between-study variance, we aim to facilitate a visual assessment of their relationship. This provides a valuable guide for practitioners to adjust traditional methods, tending to underestimate variance when the number of studies is small.

To address the challenge of simultaneously estimating the overall treatment effect and between-study variance when the number of studies is small, we make use of the flexibility of Huber's M-estimation. The M-estimation extends the principles of maximum likelihood estimation and encompasses many estimation procedures, including the least squares and least absolute deviations methods \citep{huber1992robust}. By selecting an appropriate convex function tailored to the characteristics of the data, the M-estimator more effectively captures the underlying variability in the between-study variance, leading to more accurate meta-analytic inferences. Our approach represents an improvement over traditional moment estimations by incorporating the uncertainty of the between-study variance into the inference process. This is especially useful in settings with a small number of studies, where traditional methods often yield overly narrow and unreliable CIs.

To validate the performance of our proposed method, we conducted simulations and applied it to a practical meta-analysis involving patients with juvenile idiopathic arthritis (JIA). Since the between-study variance was estimated to be zero in the example, existing methods estimate the CI of the overall treatment effect as if there were no between-study heterogeneity. Our method allows for a visual assessment of the relationship between the overall treatment effect and the between-study variance while maintaining accuracy in the interval estimation of the treatment effect. Furthermore, the method proposed in this paper has been implemented in the \texttt{rma.Muni()} function of the R package \texttt{metaMest}, available on GitHub, making it readily accessible for practical use. This package enables researchers to easily apply the method to their own meta-analyses, which in turn promotes transparency and reproducibility in statistical practice.

The remainder of this paper is structured as follows. Section 2 reviews traditional random-effects meta-analysis methods and introduces our new estimation approach for the true between-study variance. Section 3 discusses the theoretical properties of the proposed method, including the derivation of distributions for generalized moment estimators and the Sidik-Jonkman (SJ) estimator, as well as extensions to conservative estimation approaches. Section 4 presents the results of our simulation studies and applies the proposed method to a meta-analysis of actual data involving JIA patients. Finally, Section 5 concludes the paper by highlighting the key findings and implications for future research.

\section{Methods}\label{sec2}
We discuss typical approaches to random-effects meta-analysis and propose a novel inference method for the overall treatment effect based on Huber's M-estimation, employing bias-robust estimators. The term 'bias-robust' is used here as defined by \cite{rousseeuw1982most}, referring to estimators that minimize bias even in the presence of small changes or outliers in the meta-analyzed studies.

\subsection{Random-effects meta-analysis}
We perform a meta-analysis using $K$ studies. The estimated treatment effect of the $k$-th study, $\hat{\theta}_k$, is assumed as
\begin{align*}
    \hat{\theta}_k \mid \theta_k, \sigma_k^2 \sim N(\theta_k, \sigma_k^2),
\end{align*}
where $\theta_k$ and $\sigma_k^2$ represent the treatment effect and the within-study variance in the $k$-th study, respectively. The within-study variances, $\sigma_k^2$, are assumed to be known under the assumption that the within-study sample sizes are sufficiently large, allowing the normality approximation to hold. We consider this assumption to be valid throughout this paper.

In random-effects meta-analysis, the effect of the $k$-th study, $\theta_k$, is assumed to follow the distribution
\begin{align*}
    \theta_k \mid \theta, \tau^{*2} \sim N(\theta, \tau^{*2}),
\end{align*}
where $\theta$ and $\tau^{*2}$ represent the true overall treatment effect and between-study variance, respectively. By marginalizing over $\theta_k$, the model becomes
\begin{align}
\label{eq:rma}
    \hat{\theta}_k \mid \theta, \tau^{*2} \sim N(\theta, \sigma_k^2 + \tau^{*2}).
\end{align}
The model \eqref{eq:rma} is used to infer the overall treatment effect $\theta$, with the between-study variance $\tau^{*2}$ representing the variability in effects between studies and influencing the estimation of $\theta$.

Under known $\tau^{*2}$, the overall treatment effect can be estimated as
\begin{align}
\label{eq:hattheta}
    \hat{\theta}(\tau^{*2}) &= \frac{\sum_{k=1}^{K} w_k(\tau^{*2}) \hat{\theta}_k}{\sum_{k=1}^{K} w_k(\tau^{*2})},
\end{align}
where $w_k(\tau^{*2})=(\sigma_k^2+\tau^{*2})^{-1}$.
The expectation and variance of this estimator are given by $E[\hat{\theta}(\tau^{*2})]=\theta$ and $V[\hat{\theta}(\tau^{*2})] = \left( \sum_{k=1}^K w_k(\tau^{*2}) \right)^{-1}$, respectively.
Thus, the estimator $\hat{\theta}(\tau^{*2})$ is unbiased and consistent.
The test statistic $T(\tau^{*2}) = (\hat{\theta}(\tau^{*2})-\theta_0)/\sqrt{V[\hat{\theta}]}$ follows a standard normal distribution under the null hypothesis $\theta=\theta_0$.
The $100(1-\alpha)\%$ CI is given by
\begin{align}
\label{eq:ci-theta0}
    \hat{\theta}(\tau^{*2}) \pm z_{\alpha/2} \left(\sum_{k=1}^K w_k(\tau^{*2}) \right)^{-1},
\end{align}
where $z_{\alpha/2}$ is the $100(1-\alpha/2)\%$ quantile of the standard normal distribution. However, $\tau^{*2}$ is usually unknown in real data analysis, so a between-study variance estimator $\hat{\tau}^2$ is often substituted for $\tau^{*2}$ in \eqref{eq:hattheta} and \eqref{eq:ci-theta0}.
Many authors have discussed methods for estimating $\tau^{*2}$. For instance, \citet{dersimonian1986meta} and \citet{sidik_comparison_2007} propose moment-based estimators, while \citet{raudenbush2009analyzing} suggests restricted maximum likelihood estimation. Additionally, \citet{veroniki2016methods} provides a comprehensive comparison of these and other methods, highlighting their strengths and limitations across different meta-analysis scenarios.

\subsection{Distribution of estimated overall treatment effect}
Although the CI in \eqref{eq:ci-theta0} with the true between-study variance $\tau^{*2}$ achieves nominal coverage probability, the CI in \eqref{eq:ci-theta0}, where the unknown $\tau^{*2}$ is replaced by its estimate $\hat{\tau}^2$, may not reach the nominal level because $\hat{\tau}^2$ is treated as a constant in spite of being a random variable.
We consider the marginal distribution of the overall treatment effect with the between-study variance estimator using the M-estimator framework and propose a CI and $p$-value based on this marginal distribution.

First, we decompose the simultaneous density function of the pair of estimators $(\hat{\theta}, \hat{\tau}^2)$ into
\begin{align*}
    f_{\hat{\theta}, \hat{\tau}^2}(y,x\mid \theta, \tau^{*2}) &= f_{\hat{\theta}\mid \hat{\tau}^2}(y\mid x, \theta) f_{\hat{\tau}^2}(x\mid  \tau^{*2}),
\end{align*}
where the conditional density of $\hat{\theta}$ given $\hat{\tau}^2=x$, $f_{\hat{\theta}\mid \hat{\tau}^2}(y\mid x, \theta)$ is always normal with mean $\theta$ and variance $1/\sum_{k=1}^K w_k(x)$. The density function $f_{\hat{\tau}^2}(\cdot\mid \tau^{*2})$ of $\hat{\tau}^2$ varies according to the estimation method of between-study variance and depends on the true parameter $\tau^{*2}$.

The distribution function of the overall treatment effect is obtained by marginalizing over the between-study variance:
\begin{align*}
    F_{\hat{\theta}}(y\mid \tau^{*2}) &= \int_{-\infty}^{y} \int_{-\infty}^{\infty} f_{\hat{\theta}\mid \hat{\tau}^2} (z\mid x) f_{\hat{\tau}^2}(x\mid \tau^{*2}) ~dx dz \\
    &= \int_{-\infty}^{\infty} F_{\hat{\theta}\mid \hat{\tau}^2} (y\mid x) f_{\hat{\tau}^2}(x\mid \tau^{*2}) ~dx.
\end{align*}
The formulations $f_{\hat{\theta}, \hat{\tau}^2}$ and $F_{\hat{\theta}}$ do not assume that $\hat{\tau}^2$ equals the true between-study variance $\tau^{*2}$, but instead integrate over the uncertainty in $\hat{\tau}^2$ through marginalization. The second factor, $f_{\hat{\tau}^2}(x\mid \tau^{*2})$, represents the distribution of the estimated variance, capturing the variability in $\tau^{*2}$. By marginalizing across the two density functions, we account for the uncertainty in $\tau^{*2}$ and reflect this uncertainty in the distribution of the overall treatment effect.

\subsection{New estimation for the true between-study variance $\tau^{*2}$}
Suppose that the true between-study variance $\tau^{*2}$ is unknown. We propose two estimation procedures for between-study variance $\tau^2$ applying Huber's M-estimation. First, we introduce the concept of the M-estimation.
Let us consider constructing the between-study variance estimator $\hat{\tau}^2$ using the M-estimation from the density function $f_{\hat{\tau}^2}(\cdot\mid \tau^2)$. The goal is to determine how estimated between-study variance should be corrected under a given $\hat{\tau}^2$. 
In our context, the M-estimator as defined by \citet{huber1992robust} is an argmin-solution $\nu$ provided by 
\begin{align}\label{huber-M}
    \underset{\nu \ge 0}{\mathrm{argmin}}~E[ \rho(\hat{\tau}^2, \nu) \mid  \tau^2 ],
\end{align}
where $$E[ \rho(\hat{\tau}^2, \nu) \mid  \tau^2 ] = \int_{0}^{\infty} \rho(x, \nu) f_{\hat{\tau}^2} (x\mid  \tau^2) ~dx,$$ $\rho(s,v)$ is a real-valued convex function. We can choose the $\rho(s,v)$ according to the estimand. For example, when $\rho(s,\tau^2) = -\frac{d}{d \tau^2} \log f_{\hat{\tau}^2}(s\mid \tau^2)$, the solution of \eqref{huber-M} maximizes the probability density function of the estimated between-study variance with respect to $\tau^2$. This is equivalent to maximizing Fisher's fiducial distribution or the confidence density distribution of $\hat{\tau}^2$ \citep{fisher1935fiducial, singh2007confidence}.
For another example, when $\tau^{*2}$ is the scale parameter, the solution of \eqref{huber-M} with $\rho(s,v) = (\log s - \log v)^2$ corresponds the least squares estimator, while the solution of \eqref{huber-M} with $\rho(s,v) = | \log s - \log v | $ minimize the absolute deviations. Alternatively, when $\rho(s,v) = (s-v)^2$ or $\rho(s,v) = | s-v |$, the solution of \eqref{huber-M} can estimate $\tau^{*2}$ as the location parameter. These examples illustrate that $\rho(s,v)$ can be selected based on the characteristics of the parameter and the desired properties of the estimator.

In the first case, suppose that $\hat{\tau}^2$ would be estimated as the M-estimator. Define the first new estimator of between-study variance $\hat{\tau}_{M_1}^2$ by
\begin{align}
\label{eq:est1}
    \hat{\tau}_{M_1}^2(\rho) &= \left\{ \tau^2 \in [0, \infty) ~\middle |~  \hat{\tau}^2 = \underset{\nu \ge 0}{\mathrm{argmin}}~\int_{0}^{\infty} \rho(x, \nu) f_{\hat{\tau}^2} (x\mid  \tau^2) ~dx \right\}.
\end{align}
A key advantage of \eqref{eq:est1} is its ability to yield the other estimator $\hat{\tau}_{M_1}^2$, which is crucial for constructing the density function of $\hat{\tau}^2$ in various methods, such as the DL estimator. Although conventional approaches have no choice but replacing the parameter $\tau^2$ in the density function with its estimate $\hat{\tau}^2$, the M-estimation can provide a more suitable alternative, $\hat{\tau}_{M_1}^2$. This makes the M-estimator a valuable extension of traditional moment-based estimation methods. We focus on the function $\rho(s,v)= |s-v| $ because it is the most bias-robust estimator \citep{rousseeuw1982most}. Thus, we have the following Proposition \ref{prop1}, which shows that $\hat{\tau}_{M_1}^2$ becomes an estimator such that  $\hat{\tau}^2$ is the median of the distribution $F_{\hat{\tau}^2}(\cdot \mid \hat{\tau}_{M_1}^2)$.

\begin{prop}\label{prop1}
Under the random-effects model \eqref{eq:rma} and the estimation procedure \eqref{eq:est1}, with $\rho(s,v) = | u - v | $, we have
\begin{align*}
    \hat{\tau}_{M_1}^2 &= \left\{ \tau^2 \in [0, \infty) ~\middle|~  F_{\hat{\tau}^2}(\hat{\tau}^2 \mid \tau^2) = \frac{1}{2} \right\}.
\end{align*}
\end{prop}
The proof of Proposition \ref{prop1} is provided in \ref{ap-M}. From Proposition \ref{prop1}, we see that the estimator $\hat{\tau}_{M_1}^2$ satisfies the conditions of Theorem 1, which will be presented in Section \ref{sec:theory}.

In the second case, suppose that $\hat{\tau}^2$ is not necessarily the M-estimator. Define the second new estimator $\hat{\tau}_{M_2}^2(\rho)$ for $\tau^{*2}$ by
\begin{align}
\label{eq:tau2m}
    \hat{\tau}_{M_2}^2(\rho) &= \underset{\tau^2 \ge 0}{\mathrm{argmin}}~\int_{0}^{\infty} \rho(x, \hat{\tau}^2) f_{\hat{\tau}^2} (x\mid \tau^2) ~dx.
\end{align}
This is the M-estimator of the between-study variance.
Under the random-effects model \eqref{eq:rma}, $\tau^2$ can be regarded as a scale parameter. The logarithmic absolute function $\rho(s,v) = | \log s - \log v | $ can be applied to estimate this scale parameter following the principles of M-estimation \citep{thall1979huber}. 
It is implicitly assumed that the values of $x$ and $\hat{\tau}^2$ in $\rho(x, \hat{\tau}^2)$ are non-zero. When $x$ or $\hat{\tau}^2$ is close to zero, a small value (e.g., $10^{-10}$) can be added to avoid numerical instability or blowup in the computation of \eqref{eq:tau2m}. For a generalized discussion on the between-study variance estimator, we denote both $\hat{\tau}_{M_1}^2$ and $\hat{\tau}_{M_2}^2$ collectively as $\hat{\tau}_M^2$ regardless of the two estimation procedures.

Two propositions for $\hat{\tau}_M^2$, with detailed statements and proofs provided as Proposition \ref{ap-M:prop1} and Proposition \ref{ap-M:prop2} in \ref{ap-M}, state that (i) the variance of $\hat{\tau}_M^2$ varies by the estimation procedure of $\hat{\tau}^2$ including the true heterogeneity $\tau^{*2}$ and (ii) if $\hat{\tau}^2$ is consistent with $\tau^{*2}$, then $\hat{\tau}_M^2$ is also consistent.

The distribution function of the overall treatment effect, $F_{\hat{\theta}} (y\mid \tau^{*2})$, can be estimated as
\begin{align}
\label{eq:cdf}
    \hat{F}_{\hat{\theta}} (y\mid \hat{\tau}_M^2) &= \int_{-\infty}^{\infty} F_{\hat{\theta}\mid \hat{\tau}^2} (y\mid x) f_{\hat{\tau}^2}(x\mid \hat{\tau}_M^2) ~dx.
\end{align}
The following proposition provides a rationale for estimating the distribution function using $\hat{\tau}_M^2$. Especially since the estimated distribution function converges uniformly on $y$ in probability to the exact distribution as the number of studies goes to infinity, theoretical validity in the interval estimation and testing for $\theta$ is guaranteed.

\begin{prop}\label{prop2}
Assume that under the random-effects model \eqref{eq:rma}, $\frac{\partial}{\partial \tau^2} f_{\hat{\tau}^2}(x\mid \tau^2)$ exists and is absolutely integrable. If $\hat{\tau}^2 \to_p \tau^2$, then for any real value $y$, the estimator $\hat{F}_{\hat{\theta}}(y\mid \hat{\tau}_M^2)$ converges uniformly on $y$ in probability to $F_{\hat{\theta}}(y\mid \tau^{*2})$ as $K \to \infty$, where the notation $\to_p$ denotes convergence in probability. 
\end{prop}
The proof of Proposition \ref{prop2} is provided in \ref{ap-M}. By Proposition \ref{prop2}, the $100(1-\alpha)\%$ CI of the overall treatment effect can be approximately constructed as 
\begin{align*}
    \left( \hat{F}_{\hat{\theta}}^{-1} \left(\frac{\alpha}{2} ~\middle|~ \hat{\tau}_M^2 \right), \hat{F}_{\hat{\theta}}^{-1}\left(1-\frac{\alpha}{2} ~\middle|~ \hat{\tau}_M^2\right) \right).
\end{align*}
The one- and two-sided $p$-values for the null hypothesis $H_0: \theta=\theta_0$ can be derived from the distribution function \eqref{eq:cdf} as $1-\hat{F}_{\hat{\theta}}(| \hat{\theta}(\hat{\tau}^2)-\theta_0|  \mid \hat{\tau}_M^2)$ and $2(1-\hat{F}_{\hat{\theta}}(| \hat{\theta}(\hat{\tau}^2)-\theta_0| \mid \hat{\tau}_M^2))$, respectively.

\section{Theoretical aspects of the M-estimator approach}
\label{sec:theory}
In Section \ref{sec:prop-M}, we present a main property of the M-estimator approach. Similarly, in Section \ref{sec:dist-tau2}, we investigate some estimators of the between-study variance and discuss their exact distributions. In parallel, we apply the M-estimator approach to two estimation procedures for more accurate estimation of the overall treatment effect in Section \ref{sec:HKSJ-expansion}. 

\subsection{Conservative property of the M-estimator approach}
\label{sec:prop-M}
We present the main result concerning the CI of the overall treatment effect based on the M-estimator approach. The result indicates that the M-estimator approach with the estimated distribution function consistently provides more conservative interval estimates compared to the traditional approach of directly substituting the estimated between-study variance $\hat{\tau}^2$. Moreover, the confidence level asymptotically converges to the true level as the number of studies increases.

\begin{theorem}\label{theorem1}
Suppose that $E[\hat{\tau}^2\mid \hat{\tau}_M^2(\rho)] \ge \hat{\tau}^2$ under the random-effects model \eqref{eq:rma}. Then, we have
\begin{align*}
    F_{\hat{\theta}\mid \hat{\tau}^2}(q\mid \hat{\tau}^2) &> \hat{F}_{\hat{\theta}} (q\mid \hat{\tau}_M^2), \quad \mathrm{if~} q > \theta, \\
    F_{\hat{\theta}\mid \hat{\tau}^2}(q\mid \hat{\tau}^2) &< \hat{F}_{\hat{\theta}} (q\mid \hat{\tau}_M^2), \quad \mathrm{if~} q < \theta, \\
    F_{\hat{\theta}\mid \hat{\tau}^2}(q\mid \hat{\tau}^2) &= \hat{F}_{\hat{\theta}} (q\mid \hat{\tau}_M^2), \quad \mathrm{if~} q = \theta.
\end{align*}
Let $\hat{\theta}_L$ and $\hat{\theta}_U$ be the $100\alpha/2\%$ and $100(1-\alpha/2)\%$ points of $F_{\hat{\theta}}(\cdot\mid \hat{\tau}_M^2)$, respectively.
If $\hat{\tau}^2 \to_p \tau^{*2}$ as $K\to\infty$, then we have the following:
\begin{align*}
    F_{\hat{\theta}\mid \hat{\tau}^2}(\hat{\theta}_L\mid \hat{\tau}^2) \le \hat{F}_{\hat{\theta}}(\hat{\theta}_L\mid \hat{\tau}_M^2) &\to_p \frac{\alpha}{2}, \quad \mathrm{as~} K\to\infty, \\
    F_{\hat{\theta}\mid \hat{\tau}^2}(\hat{\theta}_U\mid \hat{\tau}^2) \ge \hat{F}_{\hat{\theta}}(\hat{\theta}_U\mid \hat{\tau}_M^2) &\to_p 1 - \frac{\alpha}{2}, \quad \mathrm{as~} K\to\infty.
\end{align*}
\end{theorem}
The proof of Theorem \ref{theorem1} is provided in \ref{ap-M:proof3}. Theorem \ref{theorem1} implies that if the convex function is chosen to satisfy the given conditions, the M-estimator approach can yield a more conservative CI than the corresponding usual approach, converging to the nominal confidence level as the number of studies increases.

The proposed estimation procedure \eqref{eq:est1} with $\rho(s,v)=| s-v | $ and procedure \eqref{eq:tau2m} with $\rho(s,v)=| \log s - \log v | $ satisfy the conditions of Theorem 1 when $\hat{\tau}^2$ follows a right-skewed distribution. Most methods, including the DL and SJ methods, estimate the between-study variance in the form of a sum of squares, which typically results in the estimated between-study variance following a right-skewed distribution. In Section \ref{sec:dist-tau2}, we discuss the distribution of the estimated between-study variance in detail.

\subsection{Exact distribution of between-study variance estimator $\hat{\tau}^2$}
\label{sec:dist-tau2}
We first illustrate the distribution of the generalized moment estimator for the between-study variance, as the probability density function is required for our proposed M-estimator. Subsequently, we examine the distributions of the DL and SJ methods, which are specific cases of the generalized moment estimator.

The generalized moment estimator for the between-study variance can be obtained as follows \citep{veroniki2016methods}:
\begin{align*}
    \hat{\tau}_{G} &= \max\{ 0, \hat{\tau}_{G,u} \},
\end{align*}
with the untruncated estimator 
\begin{align*}
    \hat{\tau}_{G,u} &= \frac{Q_a - \left( \sum_{k=1}^K a_k \sigma_k^2 - A_1^{-1} \sum_{k=1}^K a_k^2 \sigma_k^2 \right)}{A_1 - A_2/A_1},
\end{align*}
where $Q_a = \sum_{k=1}^K a_k (\hat{\theta}_k-\hat{\theta}_a)^2$ is the generalized Cochran between-study variance statistic, $a_k$ represents the weights assigned to each study, $A_r=\sum_{k=1}^K a_k^r$, and $\hat{\theta}_a=A_1^{-1} \sum_{k=1}^K a_k \hat{\theta}_k$. The distribution of the generalized moment estimator can be derived as stated in the following Theorem:
\begin{theorem}\label{theorem2}
Under the random-effects model \eqref{eq:rma}, the exact distribution of $\hat{\tau}_{G,u}$ is given by
\begin{align*}
    \hat{\tau}_{G,u} \sim \frac{\sum_{r=1}^R \lambda_r \chi_{r(1)}^2 - \left( \sum_{k=1}^K a_k \sigma_k^2 - A_1^{-1} \sum_{k=1}^K a_k^2 \sigma_k^2 \right)}{A_1 - A_2/A_1},
\end{align*}
where $\lambda_1, \cdots, \lambda_R$ are the non-negative eigenvalues of the matrix $VW$, $V$ is the variance-covariance matrix of $(\hat{\theta}_1-\hat{\theta}_a, \cdots, \hat{\theta}_K-\hat{\theta}_a)$, and the element $(k,k')$ of $V$ is given by
\begin{align}
    \label{eq:vij}
    V_{kk'} &= \left\{
    \begin{array}{lc}
        \left(\sigma_k^2 + \frac{1}{A_1^2}\sum_{i=1}^K a_i^2 \sigma_i^2 - \frac{2 a_k \sigma_k^2}{A_1} \right) + \left(1 + \frac{A_2}{A_1^2} - \frac{2a_k}{A_1} \right) \tau^2, & \mathrm{if~} k=k' \\
        \left( -\frac{a_k \sigma_k^2 + a_{k'}\sigma_{k'}^2}{A_1} + \frac{1}{A_1^2}\sum_{i=1}^K a_i^2 \sigma_i^2 \right) + \left(\frac{A_2}{A_1^2} - \frac{a_k+a_{k'}}{A_1} \right)\tau^2, & \mathrm{if~} k \ne k'
    \end{array}\right.
\end{align}
$W$ is a diagonal matrix with $a_1, \cdots, a_K$, and $\chi_{1(1)}^2, \cdots, \chi_{R(1)}^2$ are independently and identically distributed chi-square variables with 1 degree of freedom.
Furthermore, the probability density function of $\hat{\tau}_{G}^2$ is given as
\begin{align*}
    f_{\hat{\tau}_{G}^2}(x) = \left\{ \begin{array}{cc}
        0, & \mathrm{if~} x < 0 \\
        F_{\hat{\tau}_{G,u}^2}(0) & \mathrm{if~} x = 0 \\
        f_{\hat{\tau}_{G,u}^2}(x) & \mathrm{if~} x > 0
    \end{array}.\right.
\end{align*}    
\end{theorem}
The proof of Theorem \ref{theorem2} is provided in \ref{app-gmm}.
The most popular DL estimator is a special case of the generalized moment estimator with weights $a_k=1/\sigma_k^2$.
The between-study variance estimator is given as
\begin{align*}
    \hat{\tau}_{DL} &= \max(0, \hat{\tau}_u^2), \quad \hat{\tau}_u^2 = \frac{Q - (K-1)}{W_1(0) - W_2(0)/W_1(0)}, \quad Q = \sum_{k=1}^K w_k(0) (\hat{\theta}_k - \hat{\theta}(0))^2, \nonumber
\end{align*}
where $W_r(\tau^2)=\sum_{k=1}^K w_k^r(\tau^2)$.
The element $(k,k')$ of the variance-covariance matrix of $(\hat{\theta}_1-\hat{\theta}(0), \cdots, \hat{\theta}_K-\hat{\theta}(0))$ is given by
\begin{align*}
    V_{kk'} &= \left\{
    \begin{array}{lc}
        \left(\sigma_k^2 - \frac{1}{W_1(0)} \right) + \left(1 + \frac{W_2(0)}{W_1^2(0)} - \frac{2}{\sigma_k^2 W_1(0)} \right) \tau^{*2}, & \mathrm{if~} k=k' \\
        -\frac{1}{W_1(0)} + \left(\frac{W_2(0)}{W_1^2(0)} - \frac{1/\sigma_k^2 + 1/\sigma_{k'}^2}{W_1(0)} \right)\tau^{*2}, & \mathrm{if~} k \ne k'
    \end{array}\right..
\end{align*}
Thus, we can construct a CI for the between-study variance from the probability distribution function as 
\begin{align*}
    \left( F_{\hat{\tau}_{DL}^2}^{-1}\left(\frac{\alpha}{2} ~\middle|~  \tau^{*2} \right), F_{\hat{\tau}_{DL}^2}^{-1}\left(1-\frac{\alpha}{2} ~\middle|~  \tau^{*2} \right) \right),
\end{align*}
where the unknown parameter $\tau^{*2}$ is estimated using the M-estimator $\hat{\tau}_M^2$ in place of $\tau^{*2}$.

Our interval construction shares similarities on the approach with the ``Q-profile" CI described by \citet{viechtbauer2007confidence}, as it also utilizes the exact distribution of the estimator. However, unlike the Q-profile method, which estimates its CI based on the distribution of Cochran's Q-statistic, our approach derives a CI specific to each generalized moment estimator. Since the CI is constructed in the framework of the generalized moment estimator, our method has integrity corresponding to many generalized moment-based estimations for the overall treatment effect. This approach is closely related to the concept of a confidence distribution, as it leverages the distribution of estimators to infer the between-study variance. We apply this approach to the DL estimator and extend it to other moment-based estimations, making it a versatile framework for constructing CIs in various meta-analysis settings.

The SJ method is another moment estimation considered within a framework different from the generalized moment estimation. The SJ between-study variance estimator is given by
\begin{align*}
    \hat{\tau}_{SJ}^2(\tau^2) &= \frac{1}{K-1} \sum_{k=1}^K v_k^{-1}(\tau^2) (\hat{\theta}_k - \bar{\theta}_v)^2,
\end{align*} 
where
\begin{align*}
\bar{\theta}_v &= \frac{\sum_{k=1}^K v_k^{-1} \hat{\theta}_k}{\sum_{k=1}^K v_k^{-1}}, \quad v_k(\tau^2) = 1 + \frac{\sigma_k^2}{\tau^{2}}.
\end{align*}
Typically, $\hat{v}_k = 1 + K \sigma_k^2/\sum_{k=1}^K (\hat{\theta}_k - \bar{\theta})^2$ is used instead of $v_k(\tau^2)$, where $\bar{\theta} = \frac{1}{K} \sum_{k=1}^K \hat{\theta}_k$.
In our approach, the SJ estimator with an unknown parameter $\tau^2$ can be estimated as the solution of estimation procedure \eqref{eq:est1} or \eqref{eq:tau2m}, using $\hat{\tau}_{SJ}^2(\tau^2)$ instead of the fixed estimate $\hat{\tau}^2$, which does not depend on $\tau^2$.
We present the distribution for the SJ estimator of the between-study variance in the next Corollary.

\begin{cor}\label{theorem3}
Under the random-effects model \eqref{eq:rma}, the SJ estimator of between-study variance is distributed as 
\begin{align*}
    \hat{\tau}_{SJ}^2(\tau^2) \sim \frac{\tau^2}{K-1} \chi_{K-1}^2,
\end{align*}
where $\chi_{K-1}^2$ is a chi-square distribution with $K-1$ degrees of freedom.
\end{cor}
The proof of Corollary \ref{theorem3} is provided in \ref{ap-sj}. In \citet{sidik2005simple}, $(K-1)\hat{\tau}_{SJ}^2/\tau^{*2}$ is shown to be approximately distributed as $\chi_{K-1}^2$. 
The remarkable point of Corollary \ref{theorem3} is that it is not an approximation and strictly follows the weighted chi-squared distribution. Thus, we can construct a CI in a manner similar to the DL method. From Corollary \ref{theorem3}, $\tau^2$ is a scale parameter, and $\hat{\tau}_{SJ}^2$ satisfies the condition of Theorem 1 when the convex function is either absolute or logarithmic absolute. The CI for the overall treatment effect by the SJ method with the M-estimator is more conservative than that of the typical SJ method.

\subsection{Expansion to conservative estimation procedures}
\label{sec:HKSJ-expansion}
We expand two conservative estimation procedures, the HKSJ approach and the method proposed by \citet{michael_exact_2019}, to incorporate the M-estimator approach for more accurate interval estimation. 
A method has been proposed to adjust the test statistic to approximate a $t$-distribution with $K-1$ degrees of freedom, thereby improving the accuracy of both testing and interval estimation \citep{hartung1999alternative, hartung2001tests, hartung2001refined, sidik2002simple}. 
This approach is known as the HKSJ approach. For example, when adjusting the test statistic of the DL method, the HKSJ approach modifies the test statistic with null hypothesis $H_0: \theta = \theta_0$ as follows:
\begin{align*}
    T_{HKSJ} &= \frac{\sqrt{\sum_{k=1}^K w_k(\hat{\tau}_{DL}^2)} (\hat{\theta}_{DL} - \theta_0)}{\sqrt{\left\{\sum_{k=1}^K w_k(\hat{\tau}_{DL}^2)(\hat{\theta}_k-\hat{\theta}_{DL})^2 / (K-1)\right\}}}.
\end{align*}
The test statistic $T_{HKSJ}$ approximately follows a $t$-distribution with $K-1$ degrees of freedom. 
Therefore, the CI for the overall treatment effect $\theta$ can be computed as 
\begin{align*}
    \hat{\theta}_{DL} \pm t_{K-1, \alpha/2} \sqrt{\frac{\sum_{k=1}^K w_k(\hat{\tau}_{DL}^2)(\hat{\theta}_k-\hat{\theta}_{DL})^2}{(K-1)\sum_{k=1}^K w_k(\hat{\tau}_{DL}^2)}},
\end{align*}
where $t_{K-1, \alpha/2}$ denotes the $100(1-\alpha/2)\%$ quantile of the $t$-distribution with $K-1$ degrees of freedom. 
In many cases, the HKSJ approach is more accurate than the ordinary approach, such as the DL method.
However, it is known that the actual confidence level of the HKSJ approach falls below the nominal level when the number of studies is small, especially when the studies include either large or small sample sizes \citep{rover2015hartung}.

We consider applying our M-estimation to the HKSJ approach. The HKSJ approach adjusts the test statistic to seemingly exclude heterogeneity as a parameter in the distribution, ensuring that the test statistic approximately follows a $t$-distribution. However, this adjustment is an approximation, and the exact distribution of the HKSJ test statistic still includes $\tau^2$ as a parameter.
We propose the probability distribution function of the overall treatment effect estimator $\hat{\theta}$ with the HKSJ correction, while accounting for the exact distribution of the between-study variance $\tau^2$ as below:
\begin{align*}
    F_{\hat{\theta}} (y\mid  \hat{\theta}_1, \dots, \hat{\theta}_K, \hat{\tau}_M^2) &= \int_{-\infty}^{\infty} \Pr \left\{ \hat{\theta} \le y \mid \tau^2 = x \right\} f_{\hat{\tau}^2}(x \mid \hat{\tau}_M^2) ~dx \\
    &= \int_{-\infty}^{\infty} F_{Y}\left( \{y - \hat{\theta}(x)\} \left( \frac{\sum_{k=1}^K w_k(x)(\hat{\theta}_k - \hat{\theta}(x))^2 }{(K-1)\sum_{k=1}^K w_k(x)} \right)^{-\frac{1}{2}} \right) f_{\hat{\tau}^2}(x \mid \hat{\tau}_M^2) ~dx,
\end{align*}
where $F_Y$ is the cumulative distribution function of the $t$-distribution with $K-1$ degrees of freedom.

Our proposed method addresses the exclusion of the heterogeneity parameter in the HKSJ approach by incorporating the exact distribution of the heterogeneity estimator, $\hat{\tau}^2$. This enhances the performance of the HKSJ approach, particularly when the $t$-distribution approximation is inadequate, such as when the number of subjects varies across studies. By considering the exact distribution of the estimators, our method may provide more reliable CIs. Thus, we can construct a $100(1-\alpha)\%$ CI as
\begin{align*}
    \left(F_{\hat{\theta}}^{-1}\left( \frac{\alpha}{2} ~\middle|~ \hat{\theta}_1, \dots, \hat{\theta}_K, \hat{\tau}_M^2 \right),
    F_{\hat{\theta}}^{-1}\left( 1-\frac{\alpha}{2} ~\middle|~ \hat{\theta}_1, \dots, \hat{\theta}_K, \hat{\tau}_M^2 \right)
    \right).
\end{align*}
The one- and two-sided $p$-values for the null hypothesis $H_0: \theta = \theta_0$ can also be derived as $1 - F_{\hat{\theta}}(| \hat{\theta}(\hat{\tau}^2) - \theta_0|  \mid \hat{\theta}_1, \dots, \hat{\theta}_K, \hat{\tau}_M^2)$ and $2\left(1 - F_{\hat{\theta}}(| \hat{\theta}(\hat{\tau}^2) - \theta_0|  \mid \hat{\theta}_1, \dots, \hat{\theta}_K, \hat{\tau}_M^2)\right)$, respectively.

Next, we expand an exact inference method proposed by \citet{michael_exact_2019} using the M-estimation.
Michael's method first computes the lower and upper bounds of the between-study variance $[\tau_{\text{min}}^2, \tau_{\text{max}}^2]$ via interval estimation for the between-study variance that satisfies $\Pr\{\tau_{\text{min}}^2 \le \tau^2 \le \tau_{\text{max}}^2\} \ge 1-\beta$.
The coverage probability of the CI for $\theta$ is $1-\alpha-\beta \approx 1-\alpha$ for small $\beta$.
However, calculating the CI for the between-study variance may encounter failures due to numerical issues in maximum likelihood estimation. In such cases, a solution is to set the range of $\tau_{\text{min}}^2$ and $\tau_{\text{max}}^2$ sufficiently wide. However, expanding the range increases computational cost and raises the question of whether the predefined range is appropriate. Therefore, being able to automatically set a reasonable range is meaningful. We propose applying the CI for the between-study variance based on the M-estimator to its lower and upper bounds, which are given as 
\begin{align*}
    [\tau_{\text{min}}^2, \tau_{\text{max}}^2] = \left[ 0, F^{-1}_{\hat{\tau}^2}\left( 1 - \beta \mid \hat{\tau}_M^2 \right) \right].
\end{align*}
Since the CI based on the M-estimator can be computed at any set of $\{(\hat{\theta}_1, \sigma_1^2), \dots, (\hat{\theta}_K, \sigma_K^2)\}$, the CI for the overall treatment effect by Michael's method can be extended to calculate at any study treatment effects and within-study variances without prior setting of the lower and upper bounds for the between-study variance.

\section{Numerical results}
\label{sec:num-result}
We evaluate existing methods and our proposed methods through simulations and re-analysis of real data to compare their performance. First, we introduce the setup for our simulation. Next, we present the results of our simulation for the CIs of the overall treatment effect. Finally, we re-analyze a meta-analysis with a small number of studies.

\subsection{Simulation setup}
We compare the CIs of the overall treatment effect using several methods, including the DL and SJ methods, our proposed M-estimator approach, the restricted maximum likelihood (REML) method, and the profile likelihood method with Bartlett correction (PLBC) \citep{noma2011confidence}. Specifically, the M-estimator approach \eqref{eq:est1} with an absolute convex function is applied to both the DL method (mDL1) and the SJ method (mSJ1), while \eqref{eq:tau2m} with a logarithmic absolute convex function is applied to the DL method (mDL2) and the SJ method (mSJ2). Additionally, we consider the conservative approach by \citet{michael_exact_2019} (Mi) and its extension using our M-estimator approaches \eqref{eq:est1} and \eqref{eq:tau2m} (mMi1, mMi2). The DL, SJ, and REML methods are implemented using the \texttt{metafor} package \citep{viechtbauer2010conducting}, while PLBC is implemented using the \texttt{cimeta} package \citep{nagashima2021pimeta}. Results for the HKSJ approach and its extension with our M-estimation are provided in \ref{ap-hksj}.

We adopt simulation parameters from \citet{inthout2014hartung}, as they provide a comparison of meta-analysis methods under a small number of studies. The number of studies is set to range from small to medium $(K=3, \dots, 10, 15, 20)$. Four within-study variance scenarios are considered as follows:
\begin{itemize}
    \item[1)] Equal size studies $(\sigma_1^2=\cdots=\sigma_K^2=1)$.
    \item[2)] One small study, 1/10th of the size of other studies $(\sigma_1^2=0.1, \sigma_2^2=\cdots=\sigma_K^2=1)$.
    \item[3)] 50/50 small and large studies $(\sigma_1^2=\cdots=\sigma_{K/2}^2=0.1, \sigma_{K/2+1}^2=\cdots=\sigma_K^2=1)$.
    \item[4)] Varying study sizes $(\sigma_k^2=0.1 + 0.9 \frac{k-1}{K-1}, \quad k=1, \dots, K)$.
\end{itemize}
The between-study heterogeneity is set to small, medium, and large levels $(I^2=0.25, 0.50, 0.90)$. The between-study variance $\tau^2$ is calculated based on the within-study variance and between-study heterogeneity using the formula $\tau^2=K^{-1}\sum_{k=1}^K\sigma_k^2 \frac{I^2}{1-I^2}$. The overall treatment effect is set to $\theta=0$, and 10,000 evaluations are performed for each parameter setting.

\subsection{Simulation results}
\begin{figure}
    \centerline{
    \includegraphics[width=180mm]{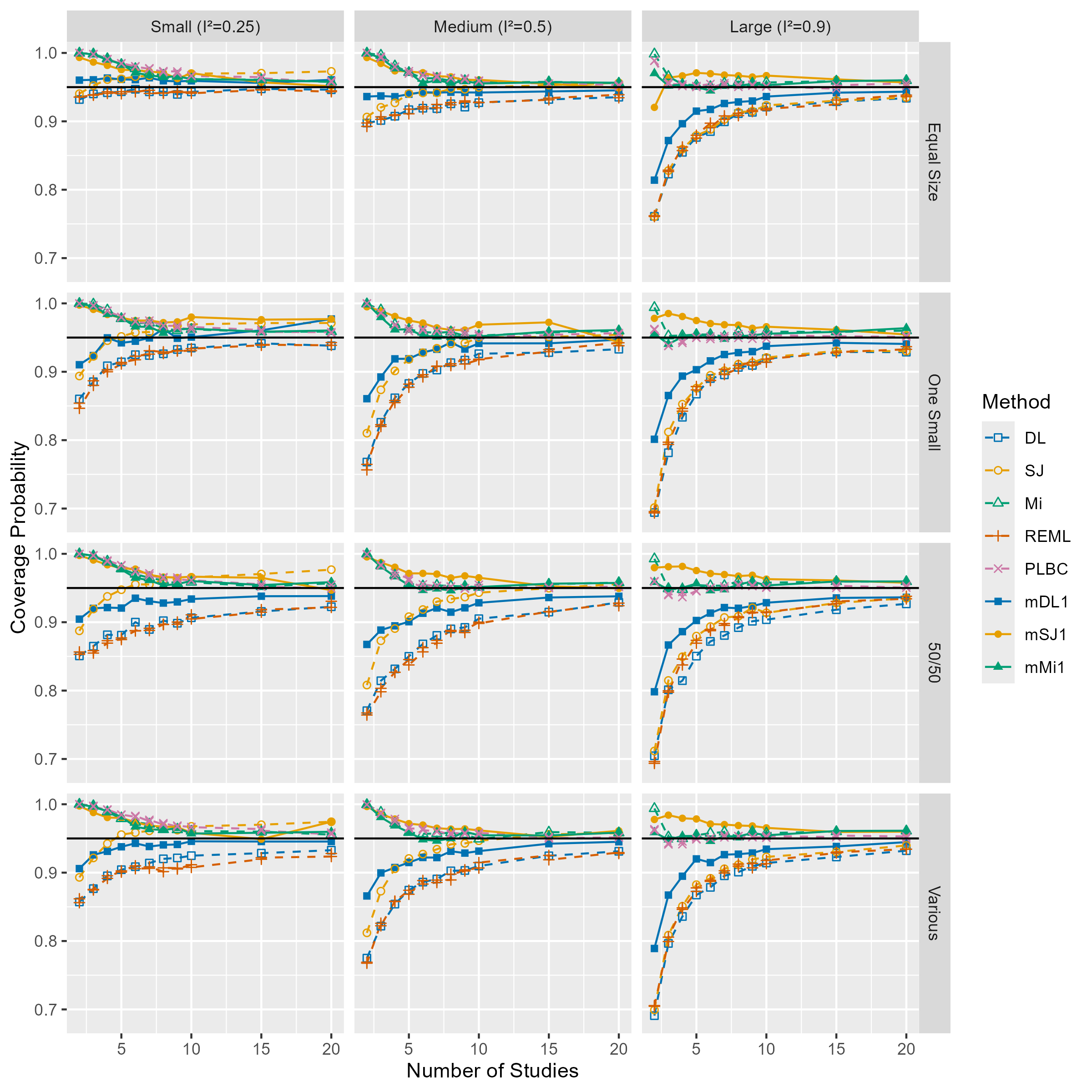}
    }
    \caption{Coverage probabilities of the CIs for the overall treatment effect using the DL, SJ, REML, PLBC, Mi, and proposed mDL1, mSJ1, and mMi1 methods. The rows correspond to the within-study variance settings, and the columns represent the between-study heterogeneity $I^2=0.25, 0.5, 0.9$.}
    \label{fig:cp-theta}
\end{figure}

\begin{figure}
    \centerline{
    \includegraphics[width=180mm]{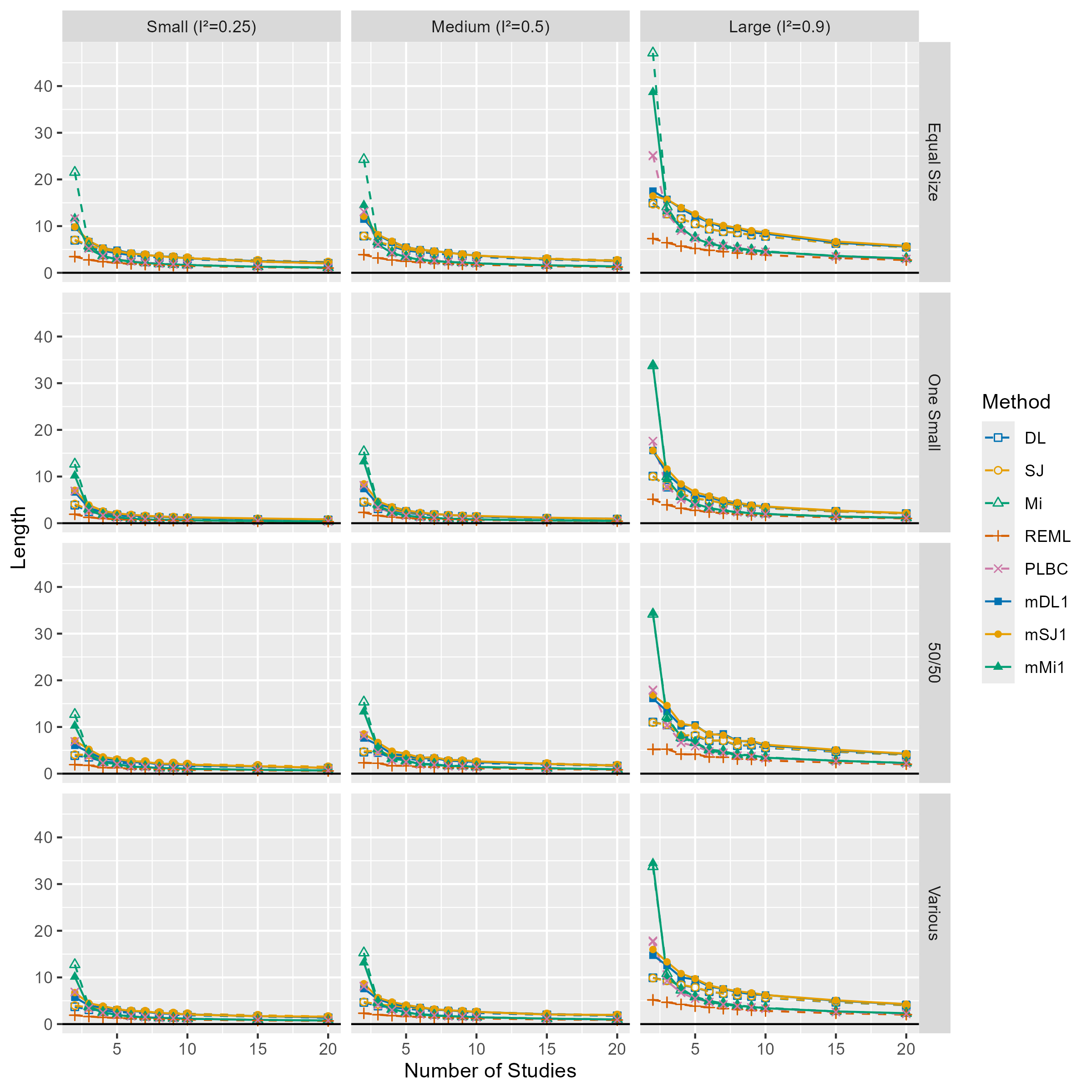}
    }
    \caption{Lengths of the CIs for the overall treatment effect using the DL, SJ, REML, PLBC, Mi, and proposed mDL1, mSJ1, and mMi1 methods. The rows correspond to the within-study variance settings, and the columns represent the between-study heterogeneity $I^2=0.25, 0.5, 0.9$.}
    \label{fig:length-theta}
\end{figure}

\begin{figure}
    \centerline{
    \includegraphics[width=180mm]{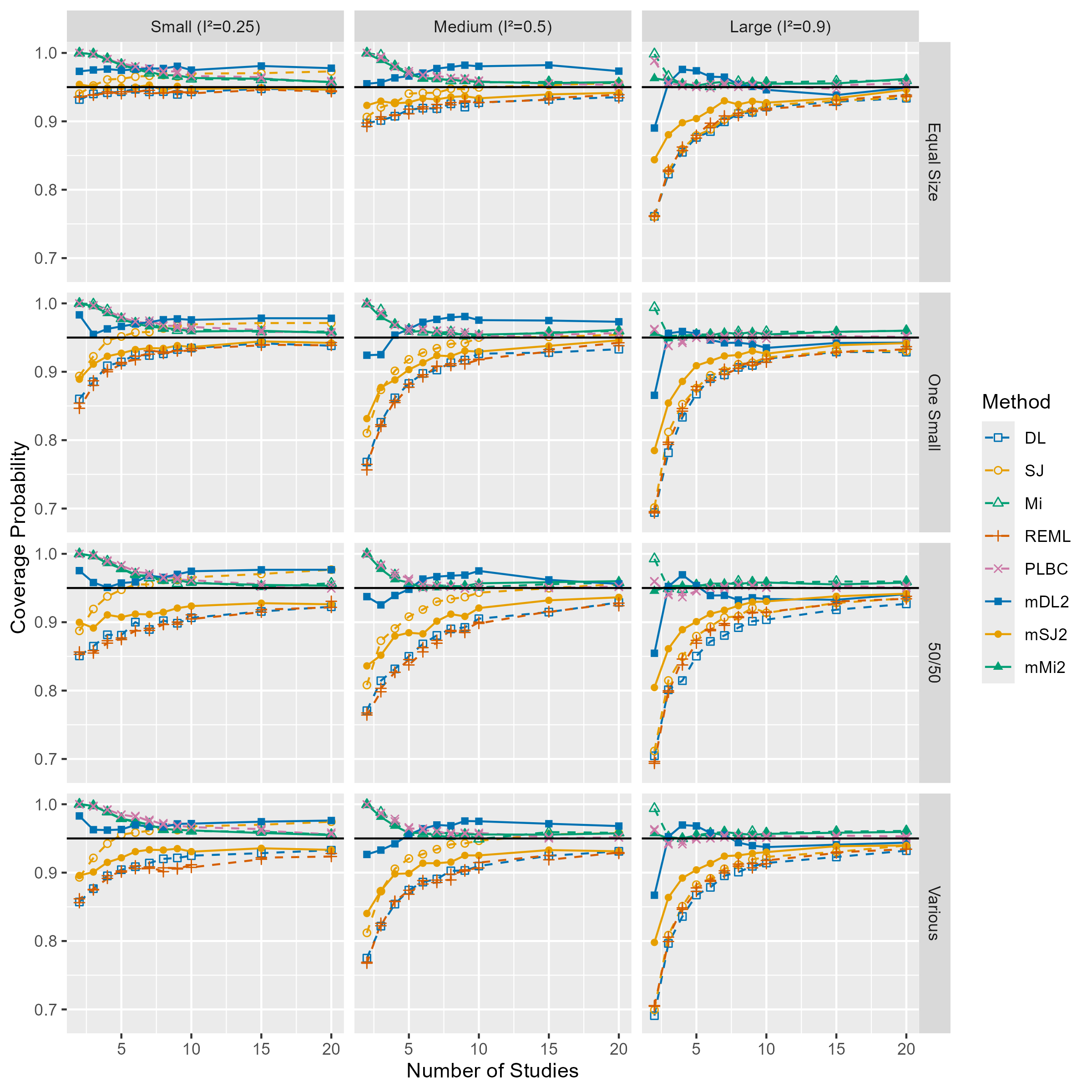}
    }
    \caption{Coverage probabilities of the CIs for the overall treatment effect using the DL, SJ, REML, PLBC, Mi, and proposed mDL2, mSJ2, and mMi2 methods. The rows correspond to the within-study variance settings, and the columns represent the between-study heterogeneity $I^2=0.25, 0.5, 0.9$.}
    \label{fig:cp-theta2}
\end{figure}

\begin{figure}
    \centerline{
    \includegraphics[width=180mm]{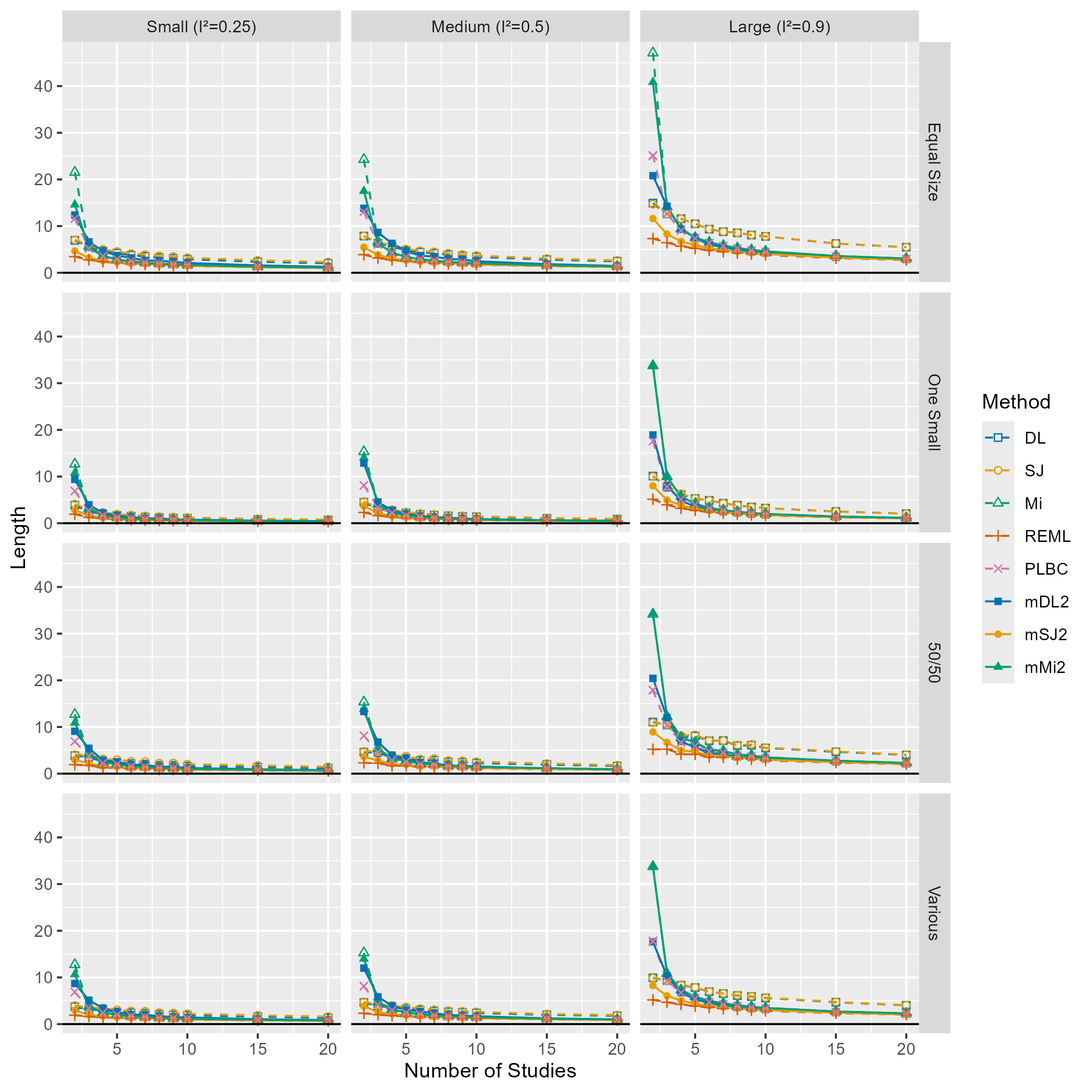}
    }
    \caption{Lengths of the CIs for the overall treatment effect using the DL, SJ, REML, PLBC, Mi, and proposed, mDL2, mSJ2, and mMi2 methods. The rows correspond to the within-study variance settings, and the columns represent the between-study heterogeneity $I^2=0.25, 0.5, 0.9$.}
    \label{fig:length-theta2}
\end{figure}

Figures \ref{fig:cp-theta}-\ref{fig:length-theta2} show the coverage probabilities and lengths of the CIs for the overall treatment effect, $\theta$, from both the comparison and our proposed methods. The M-estimator approach (mDL1, mDL2, mSJ1, mSJ2) consistently exhibited higher coverage probabilities than methods without M-estimation and REML. Greater between-study heterogeneity led to lower coverage probabilities, particularly when the number of studies was small. mDL2 and mSJ1 showed that, even with fewer than 10 studies, the coverage probability generally reached the nominal level. mSJ1 provided conservative results in almost all scenarios, even when the number of studies was small. While mDL2 tended to be overly conservative when heterogeneity was moderate or with approximately 5 studies, it remained close to the nominal level in nearly all scenarios.

In the equal-size study scenario, all methods achieved nominal level coverage probabilities when between-study heterogeneity was small. However, in other scenarios, the CIs of the DL and SJ methods narrowed, and their coverage probabilities fell below the nominal level. The Mi method was conservative, but the CI for the overall treatment effect could not be computed when the search range for the between-study variance was indeterminate.
The mMi1 and mMi2 methods were as conservative as the Mi method and had the additional advantage of being able to compute CIs without computational failures. The PLBC method also showed similarly conservative results to the Mi, mMi1, and mMi2 methods, particularly with a small number of studies, though in rare cases, it was unable to compute the CI due to issues with estimating the between-study variance. The number of computational failures is illustrated in Figure \ref{fig:N-michael} of \ref{ap-nfail}.

Notably, while the Mi method generally works even with a small number of studies, rare computational failures occur due to challenges in interval estimation of the between-study variance. Since the CI for the overall treatment effect in the Mi method depends on a predefined search range for the between-study variance, a rationale is needed for setting this range in advance. The mMi1 and mMi2 methods consistently calculated the CI, achieving nearly the same confidence level as the Mi method. In scenarios with varying within-study variance, the REML method occasionally failed due to non-convergence in the Fisher scoring algorithm, and the PLBC method also experienced rare computational failures when no estimator could be found.

\subsection{Re-analysis of data from small meta-analysis}
\begin{figure}[ht]
    \centering
    \begin{subfigure}{0.45\textwidth}
        \centering
        \includegraphics[width=\textwidth]{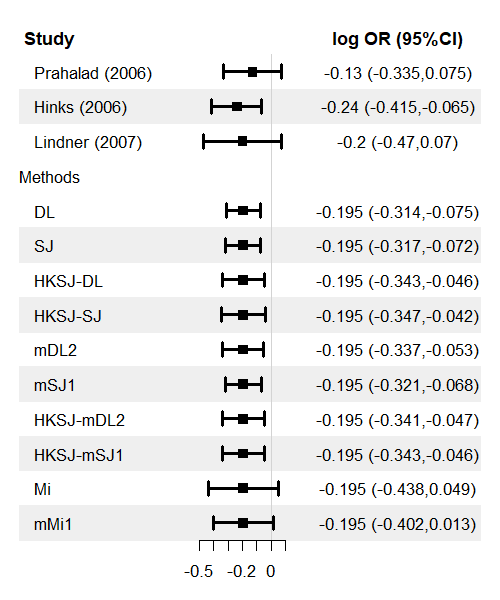}
    \end{subfigure}
    \hfill
    \begin{subfigure}{0.45\textwidth}
        \centering
        \includegraphics[width=\textwidth]{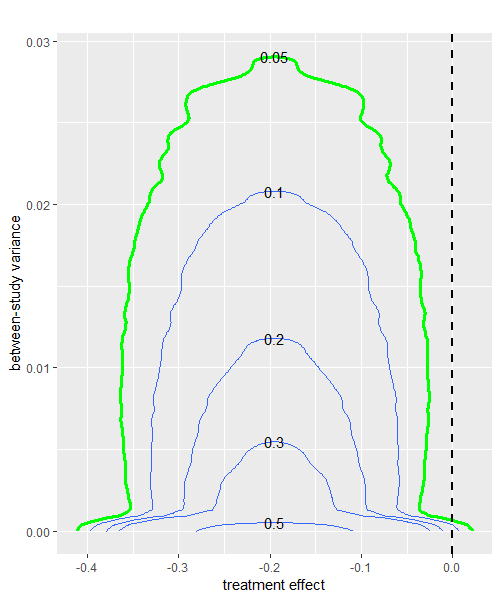}
    \end{subfigure}
    \caption{Point estimates and $95\%$ CIs of the overall treatment effect for the JIA data (Left), and the contour plot of the simultaneous density function of overall treatment effect and between-study variance from the mDL method (Right).}
    \label{fig:rda}
\end{figure}

We re-analyze the occurrence of a particular genetic variant, CCR5, in patients with juvenile idiopathic arthritis (JIA) compared to the generalized population \citep{hinks2010association}. The investigation included three studies examining the association of JIA with a biomarker via odds ratios \citep{prahalad2006association, hinks2006association, lindner2007lack}. 
We apply the DL and SJ methods, as well as our M-estimator approach stable in simulation (mDL2, mSJ1), to the log odds ratios of the treatment effect from the JIA data. The HKSJ approach for each method and Michael's method are also applied to the JIA data.
The meta-analysis results and the contour plot of the simultaneous density function from the mDL2 method, $f_{\hat{\theta}, \hat{\tau}^2}(y, x \mid \hat{\theta}_{DL}, \hat{\tau}_M^2)$, are shown in the right side of Figure \ref{fig:rda}.
The between-study variance estimated by the DL and SJ methods was zero for these data.
Compared to the typical approach, the mDL2 and mSJ1 methods produced wider CIs than the traditional DL and SJ methods and narrower CIs than the HKSJ and Mi approaches due to differences in handling the between-study variance. The Mi and mMi1 methods yielded conservative results, as the CIs included zero.

A viewpoint in our methods aids in interpreting the relationship between the estimated overall treatment effect $(\theta)$ and between-study variance $(\tau^2)$. It allows the visualization of the simultaneous density function of the estimators of $(\theta, \tau^2)$. This plot shows contour lines originating from the highest simultaneous density point, which provide probabilistic meaningful in the confidence of $(\theta, \tau^2)$. It enables us to interpret analysis results flexibly when integrating meta-analysis findings with clinical insights. For example, while the $95\%$ CI of $\theta$ does not include zero, the $95\%$ contour line of the simultaneous distribution slightly includes zero when the between-study variance is close to zero. We can use empirical, historical, and clinical knowledge to evaluate the size of between-study variation and incorporate it into our decision-making process. Furthermore, this simultaneous estimation provides a meaningful guide for considering valid hypothetical values of $(\theta, \tau^2)$ in subsequent studies. Based on the results of this meta-analysis, if a future trial assumes $\tau^2 = 0.02$, the treatment effect should be selected from the range included within the $0.05$ contour line, assuming a significance level of $0.05$. Conversely, when using individual confidence intervals for $\theta$ and $\tau^2$, appropriate adjustments to the significance level are required \citep{cupples1984multiple}. In summary, visualizing the joint confidence of $(\theta, \tau^2)$ enhances the interpretation of meta-analysis results while considering the size of between-study variance, which enables more flexible decision-making and provides valuable feedback for study design, compared to focusing solely on the treatment effect.

\section{Discussion}
We proposed a method for estimating the overall treatment effect that explicitly accounts for the distribution of the estimated between-study variance. Within the framework of traditional moment estimation, our approach provides more accurate CIs for the overall treatment effect. We identified the distribution of the between-study variance using the generalized moment method and the SJ method. The probability density function of the between-study variance, $f_{\hat{\tau}^2} (\cdot \mid \tau^2)$, includes an unknown parameter $\tau^2$. Therefore, we proposed an optimal estimator for the parameter $\tau^2$, treating the between-study variance as an M-estimator. This approach enabled us to estimate the overall treatment effect by considering the variability of the estimated between-study variance even when unknown parameters were included in the explicit distribution.
Furthermore, we illustrated the simultaneous density distribution of the overall treatment effect and between-study variance, allowing for a visual assessment of the mutual influence between these two quantities. This visualization enables researchers to better understand how variations in the between-study variance may impact the overall treatment effect.

Our simulation study demonstrated that the M-estimator approach provides more conservative estimates compared to traditional methods, such as the DL and SJ methods. A summary of method comparisons is presented in Table \ref{tab:comp-method}. In particular, the DL and SJ methods with an M-estimator yield more accurate estimates than those without an M-estimator. In supplementary simulations, the HKSJ approach combined with the M-estimator produced more precise results than the conventional HKSJ approach. The Mi method constructs conservative CIs, though it may be infeasible in rare cases due to computational failures. The mMi1 and mMi2 methods address this issue by calculating CIs without computational failures for any study treatment effects and within-study variances, and without requiring predefined bounds for between-study variance. In the re-analysis of JIA data with a small number of studies, the standard and HKSJ approaches with an M-estimator yielded CIs nearly equivalent to those obtained without an M-estimator, even when the between-study variance was zero. Our method enabled evaluation of the overall treatment effect being below zero for most values of the between-study variance by displaying a simultaneous distribution.

\begin{table}[ht]
\centering
\caption{Comparison of methods in terms of confidence level, visualization, and computational failure}\label{tab:comp-method}
\small
\begin{tabular}{| p{1.7cm}| p{3.3cm}| p{3.5cm}| p{3.5cm}| }
\hline
Method     & Confidence level with few studies & Visualization of relationship between $(\hat{\theta}, \hat{\tau}^2)$ & Computational failure \\ \hline
DL, SJ     & Lower than nominal level            & Not supported                                                             & None \\ \hline
REML       & Lower than nominal level            & Not supported                                                             & Occurs when no solution exists \\ \hline
PLBC       & Higher than nominal level           & Not supported                                                             & Occurs when no solution exists \\ \hline
Mi         & Higher than nominal level           & Not supported                                                             & Occurs when the CI for $\tau^2$ cannot be computed \\ \hline
mSJ1       & Higher than nominal level           & Visualizable via $f_{\hat{\theta}, \hat{\tau}^2}(\cdot\mid \hat{\tau}_M^2)$ & None \\ \hline
mDL2       & Higher than DL, SJ               & Visualizable via $f_{\hat{\theta}, \hat{\tau}^2}(\cdot\mid \hat{\tau}_M^2)$ & None \\ \hline
mMi1, mMi2 & Higher than nominal level           & Visualizable via $f_{\hat{\theta}, \hat{\tau}^2}(\cdot\mid \hat{\tau}_M^2)$ & None \\ \hline
\end{tabular}
\end{table}

One limitation of this study is that while we provided the distribution of the between-study variance for generalized moment estimation, we did not provide the distribution for the restricted maximum likelihood estimator due to the complexity of iterative computations. Such computations make it challenging to explicitly illustrate the distribution of the between-study variance.
Additionally, the convex function used for the M-estimator was limited to absolute value and logarithmic absolute functions, which are considered the most bias-robust estimators as shown in \citet{rousseeuw1982most}. However, it may be possible to identify more optimal convex functions depending on the distribution of the applied estimators.

In summary, our method enables a more accurate estimation of the overall treatment effect, particularly when the number of studies is small and traditional asymptotic methods are not reliable. It also facilitates visualization of the relationship between the overall treatment effect and between-study variance using the simultaneous density function, regardless of the number of studies. When the number of studies is small or when exploring the relationship between the overall treatment effect and between-study variance, we recommend using the M-estimator approach.

\section*{Acknowledgements}
This work was supported by JSPS KAKENHI (Grant numbers JP24K23862 and JP24K14853).

\section*{Supplementalry materials}
Our M-estimator approach, proposed in Section 2, and the random sampling function for DL between-study variance, described in Section 3, can be installed from GitHub\\(\texttt{https://github.com/keisuke-hanada/metaMest/}) as the \texttt{metaMest} package. The code for the re-analysis of the JIA data is included in the package example.

\appendix

\section{Proofs and two propositions}
\subsection{Proof of Theorem 1}
\label{ap-M:proof3}
We have the following identity equation:
\begin{align*}
    F_{\hat{\theta}} (q \mid \hat{\tau}_M^2) - F_{\hat{\theta} \mid \hat{\tau}^2}(q \mid \hat{\tau}^2) &= \int_0^{\infty} \left\{ F_{\hat{\theta} \mid \hat{\tau}^2}(q \mid x) - F_{\hat{\theta} \mid \hat{\tau}^2}(q \mid \hat{\tau}^2) \right\} f_{\hat{\tau}^2}(x \mid \hat{\tau}_M^2) ~dx.
\end{align*}
Now, we consider the Taylor expansion of $F_{\hat{\theta} \mid \hat{\tau}^2}(q \mid x)$ around $x = \hat{\tau}^2$. Then, there exists a constant $c$ such that
\begin{align*}
    F_{\hat{\theta}} (q \mid \hat{\tau}_M^2) - F_{\hat{\theta} \mid \hat{\tau}^2}(q \mid \hat{\tau}^2) &= \int_0^{\infty} \left. \frac{d}{~dx}F_{\hat{\theta} \mid \hat{\tau}^2} (q \mid x) \right|_{x=c} (x - \hat{\tau}^2) f_{\hat{\tau}^2}(x \mid \hat{\tau}_M^2) ~dx \\
    &= \left.\frac{d}{~dx}F_{\hat{\theta} \mid \hat{\tau}^2} (q \mid x)\right|_{x=c} \int_0^{\infty} (x - \hat{\tau}^2) f_{\hat{\tau}^2}(x \mid \hat{\tau}_M^2) ~dx \\
    &= \left.\frac{d}{~dx}F_{\hat{\theta} \mid \hat{\tau}^2} (q \mid x)\right|_{x=c} \left( E[\hat{\tau}^2 \mid \hat{\tau}_M^2] - \hat{\tau}^2 \right).
\end{align*}
Because we have $\hat{\theta} \mid \hat{\tau}^2 \sim N\left(\theta, \left( \sum_{k=1}^K w_k(x) \right)^{-1} \right)$, we can derive
\begin{align}
\label{eq:int-th}
    \frac{d}{~dx}F_{\hat{\theta} \mid \hat{\tau}^2} (q \mid x) &= \frac{d}{~dx} \int_{-\infty}^{q} \frac{1}{\sqrt{2\pi (\sum_{k=1}^K w_k(x))^{-1}}} \exp\left\{-\frac{(y-\theta)^2}{2(\sum_{k=1}^K w_k(x))^{-1}}\right\} dy \nonumber \\
    &= \int_{-\infty}^{q} \frac{d}{~dx} \frac{1}{\sqrt{2\pi (\sum_{k=1}^K w_k(x))^{-1}}} \exp\left\{-\frac{(y-\theta)^2}{2(\sum_{k=1}^K w_k(x))^{-1}}\right\} dy \nonumber \\ 
    &= \int_{-\infty}^{q} \left(1 - (y-\theta)^2 \sum_{k=1}^K w_k(x)\right) \frac{1}{\sqrt{2\pi (\sum_{k=1}^K w_k(x))^{-1}}} \nonumber \\
    &\times \sum_{k=1}^K \frac{-w_k^2(x)}{2} \exp\left\{-\frac{(y-\theta)^2}{2(\sum_{k=1}^K w_k(x))^{-1}}\right\} dy \nonumber \\
    &= -\frac{1}{2} \sum_{k=1}^K w_k^2(x) \int_{-\infty}^{(q-\theta)\sum_{k=1}^K w_k(x)} (1 - z^2) \frac{1}{\sqrt{2\pi}} e^{-\frac{z^2}{2}} dz.
\end{align}
For any real number $c_0$, we have
\begin{align}
\label{eq:pm}
    \int_{-\infty}^{c_0} (1 - z^2) \frac{1}{\sqrt{2\pi}} e^{-\frac{z^2}{2}} dz &= \int_{-\infty}^{\infty} (1 - z^2) \frac{1}{\sqrt{2\pi}} e^{-\frac{z^2}{2}} dz - \int_{c_0}^{\infty} (1 - z^2) \frac{1}{\sqrt{2\pi}} e^{-\frac{z^2}{2}} dz \nonumber \\
    &= - \int_{c_0}^{\infty} (1 - z^2) \frac{1}{\sqrt{2\pi}} e^{-\frac{z^2}{2}} dz.
\end{align}
Using the symmetry of the normal distribution and the properties of an even function, we can conclude:
\begin{align*}
    \int_{-\infty}^{0} (1 - z^2) \frac{1}{\sqrt{2\pi}} e^{-\frac{z^2}{2}} dz &= \frac{1}{2} \int_{-\infty}^{\infty} (1 - z^2) \frac{1}{\sqrt{2\pi}} e^{-\frac{z^2}{2}} dz = 0.
\end{align*}
When $q>\theta$, we have:
\begin{align*}
    \int_{-\infty}^{(q-\theta)\sum_{k=1}^K w_k(x)} (1 - z^2) \frac{1}{\sqrt{2\pi}} e^{-\frac{z^2}{2}} dz
    &= \int_{-\infty}^{0} (1 - z^2) \frac{1}{\sqrt{2\pi}} e^{-\frac{z^2}{2}} dz + \int_{0}^{(q-\theta)\sum_{k=1}^K w_k(x)} (1 - z^2) \frac{1}{\sqrt{2\pi}} e^{-\frac{z^2}{2}} dz \\
    &= \int_{0}^{1} (1 - z^2) \frac{1}{\sqrt{2\pi}} e^{-\frac{z^2}{2}} dz + \int_{1}^{(q-\theta)\sum_{k=1}^K w_k(x)} (1 - z^2) \frac{1}{\sqrt{2\pi}} e^{-\frac{z^2}{2}} dz \\
    &> \int_{0}^{1} (1 - z^2) \frac{1}{\sqrt{2\pi}} e^{-\frac{z^2}{2}} dz + \int_{1}^{\infty} (1 - z^2) \frac{1}{\sqrt{2\pi}} e^{-\frac{z^2}{2}} dz \\
    &= 0.
\end{align*}
Hence, from \eqref{eq:int-th}, we find that $\frac{d}{~dx}F_{\hat{\theta} \mid \hat{\tau}^2} (q \mid x) \mid_{x=c} < 0~(q>\theta)$.
When $q<\theta$, we also have $\frac{d}{~dx}F_{\hat{\theta} \mid \hat{\tau}^2} (q \mid x) \mid_{x=c} < 0$ based on \eqref{eq:pm}.
As a result, we obtain:
\begin{align*}
    \left.\frac{d}{~dx}F_{\hat{\theta} \mid \hat{\tau}^2} (q \mid x)\right|_{x=c} &< 0 \,\,\,\,\, (q > \theta) \\
    \left.\frac{d}{~dx}F_{\hat{\theta} \mid \hat{\tau}^2} (q \mid x)\right|_{x=c} &> 0 \,\,\,\,\, (q < \theta) \\
    \left.\frac{d}{~dx}F_{\hat{\theta} \mid \hat{\tau}^2} (q \mid x)\right|_{x=c} &= 0 \,\,\,\,\, (q = \theta).
\end{align*}
Thus, if $E[\hat{\tau}^2 \mid \hat{\tau}_M^2] \ge \hat{\tau}^2$, the $100(1-\alpha)\%$ CI of $\theta$ using $\hat{\tau}_M^2$ is more conservative than that using $\hat{\tau}^2$.
From Proposition 4 in Appendix A.2, if $\hat{\tau}^2 \to_p \tau^{*2}$, then $\hat{\tau}_M^2 \to_p \tau^{*2}$.
Thus, for any real number $q$, we have:
\begin{align*}
    \left| F_{\hat{\theta}} (q \mid \hat{\tau}_M^2) -  F_{\hat{\theta}} (q \mid \tau^{*2}) \right|
    &= \left|  \left\{\left.\frac{dF_{\hat{\theta}}}{d\tau^2}(q \mid \tau^2)\right| _{\tau^2=\tau^2_c} \right\} (\hat{\tau}_M^2 - \tau^{*2}) \right|   \\
    &\le \left|  \left\{ \left.\frac{dF_{\hat{\theta}}}{d\tau^2}(q \mid \tau^2)\right| _{\tau^2=\tau^2_c} \right\} \right|  \left|  \hat{\tau}_M^2 - \tau^{*2} \right| 
    \to_p 0 \,\,\, (K \to \infty).
\end{align*}
Therefore, the coverage probability of the $100(1-\alpha)\%$ CI for $\theta$ converges to its true value as the number of studies increases.

\rightline{$\square$}

\subsection{Some properties of M-estimator}
\label{ap-M}
\begin{prop}\label{ap-M:prop1}
Under the random-effects model \eqref{eq:rma}, the variance of the M-estimator $\hat{\tau}_M^2$ is given as
\begin{align*}
    V[\hat{\tau}_M^2] &= \left(\frac{1}{r(\tau^{*2})}\right)^2 V[\hat{\tau}^2 \mid \tau^2 = \tau^{*2}] + O_p(a_K^2) \,\,\,\,\, \mathrm{as~} K \to \infty,
\end{align*}
where $r(x) = \frac{dR}{d\tau^2}(x)$, $a_K = r(c_K) - r(\tau^{*2})$, $c_K$ is a positive constant such that $\mid c_K - \tau^{*2}\mid  < \mid \hat{\tau}_M^2 - \tau^{*2}\mid $, $R$ is the function of $\hat{\tau}_M^2$ from \eqref{eq:est1} or \eqref{eq:tau2m}, defined as follows:
\begin{align}
\label{eq:R1}
    R(\hat{\tau}_M^2) &= \underset{t \ge 0}{\mathrm{argmin}}~\int_{-\infty}^{\infty} \rho(x,t) f_{\hat{\tau}^2} (x\mid \hat{\tau}_M^2) ~dx = \hat{\tau}^2, \\
\label{eq:R2}
    R^{-1}(\hat{\tau}^2) &= \underset{\tau^2 \geq 0}{\mathrm{argmin}} \int_{0}^{\infty} \rho(x, \hat{\tau}^2) f_{\hat{\tau}^2}(x \mid \tau^2) ~dx = \hat{\tau}_M^2,
\end{align}
and $O_p(a_K)$ denotes the order less than or equal to that of $a_K$ in probability.
Thus, the variance estimator of $\hat{\tau}_M^2$ is
\begin{align*}
    \hat{V}[\hat{\tau}_M^2] &= \left(\frac{1}{r(\hat{\tau}_M^2)}\right)^2 V[\hat{\tau}^2 \mid \tau^2 = \hat{\tau}_M^2].
\end{align*}
\end{prop}
To prove Proposition 3, we first consider a Taylor expansion of $\hat{\tau}^2 = R(\hat{\tau}_M^2)$ around $\hat{\tau}_M^2 = \tau^{*2}$. In this case, there exists a positive constant $c_K > 0$ such that
\begin{align*}
    \hat{\tau}^2 &= R(\hat{\tau}_M^2) = R(\tau^{*2}) + r(c_K) (\hat{\tau}_M^2 - \tau^{*2}),
\end{align*}
where $\mid c_K - \tau^{*2}\mid  < \mid \hat{\tau}_M^2 - \tau^{*2}\mid $ and $r(c) = \frac{dR}{d\tau^2}(c)$.

Thus, the variance of $\hat{\tau}_M^2$ is written as
\begin{align*}
    V[\hat{\tau}_M^2] &= \left(\frac{1}{r(c_K)}\right)^2 V[\hat{\tau}^2] \\
    &= \left(\frac{1}{r(\tau^{*2})}\right)^2 V[\hat{\tau}^2] + \left(\frac{r(c_K) - r(\tau^{*2})}{r(\tau^{*2})}\right)^2 V[\hat{\tau}^2] \\
    &= \left(\frac{1}{r(\tau^{*2})}\right)^2 V[\hat{\tau}^2] + \left(\frac{\sqrt{V[\hat{\tau}^2]}}{r(\tau^{*2})}\right)^2 a_K^2.
\end{align*}

Since $r(x) > 0$ because $R(x)$ is a monotonic increasing function, $V[\hat{\tau}^2] < \infty$, and $c_K \to_p \tau^{*2}$, the second term on the right-hand side can be calculated as follows:
\begin{align*}
    \left(\frac{\sqrt{V[\hat{\tau}^2]}}{r(\tau^{*2})}\right)^2 a_K^2 &\leq A a_K^2 \to_p 0 \,\,\, (K \to \infty),
\end{align*}
where $A (< \infty)$ is a positive constant. Hence, we can obtain
\begin{align*}
    V[\hat{\tau}_M^2] &= \left(\frac{1}{r(\tau^{*2})}\right)^2 V[\hat{\tau}^2] + O_p(a_K^2).
\end{align*}

\rightline{$\square$}

\begin{prop}\label{ap-M:prop2}
Under the random-effects model \eqref{eq:rma}, if $\hat{\tau}^2 \to_p \tau^{*2}$ as $K \to \infty$, then $\hat{\tau}_M^2 \to_p \tau^{*2}$, as $K \to \infty$.
\end{prop}
To prove Proposition 4, assume that $\hat{\tau}^2 \to_p \tau^{*2}$. Then, for any $\varepsilon > 0$, we have
\begin{align*}
    \Pr\left\{\mid \hat{\tau}^2 - \tau^{*2}\mid  > \varepsilon\right\} \to 0 \,\,\, (K \to \infty).
\end{align*}
Now, we evaluate the difference between $\hat{\tau}_M^2$ and $\tau^{*2}$ as
\begin{align*}
    \mid \hat{\tau}_M^2 - \tau^{*2}\mid  &\leq \left|  R^{-1}(\hat{\tau}^2) - R^{-1}(\tau^{*2}) \right|  + \left|  R^{-1}(\tau^{*2}) - \tau^{*2} \right| .
\end{align*}
The inverse of the estimation function $R^{-1}(\hat{\tau}^2)$ can be evaluated as:
\begin{align*}
    \mid R^{-1}(\hat{\tau}^2) - R^{-1}(\tau^{*2})\mid  &= \left|  \left\{ \frac{dR^{-1}}{d\tau^2}(\tau^2) \right\}_{\tau^2 = c} \right|  \mid \hat{\tau}^2 - \tau^{*2}\mid  \\
    &\leq M \mid \hat{\tau}^2 - \tau^{*2}\mid  \to_p 0 \,\,\, (K \to \infty),
\end{align*}
where $M (< \infty)$ is a positive constant.

Thus, we have $R^{-1}(\hat{\tau}^2) \to_p R^{-1}(\tau^{*2}) \,\,\, (K \to \infty)$. From the consistency of $\hat{\tau}^2 \to_p \tau^{*2}$ and the dominated convergence theorem, we get $\lim_{K \to \infty} R^{-1}(\tau^{*2}) = \tau^{*2}$. Thus, the first inequality converges, $\mid \hat{\tau}_M^2 - \tau^{*2}\mid  \to_p 0 \,\,\, (K \to \infty)$, and therefore, $\hat{\tau}_M^2 \to_p \tau^{*2}$.

\rightline{$\square$}

\subsubsection{Proof of Proposition \ref{prop1}}\label{ap-prop1}
The M-estimator from \eqref{eq:est1} can be rewritten as
\begin{align*}
    \hat{\tau}^2 &= \underset{t \ge 0}{\mathrm{argmin}}~\int_{0}^{\infty} \rho(x, t) f_{\hat{\tau}^2} (x\mid  \hat{\tau}_M^2) ~dx \\
    &= \underset{t \ge 0}{\mathrm{argmin}}~\left[ \int_{t}^{\infty} (x-t) f_{\hat{\tau}^2}(x\mid \hat{\tau}_M^2) ~dx - \int_{0}^{t} (x-t) f_{\hat{\tau}^2}(x\mid \hat{\tau}_M^2) ~dx \right] \\
    &= \left\{ t \in [0,\infty) \left|  \int_{t}^{\infty} f_{\hat{\tau}^2}(x\mid \hat{\tau}_M^2) ~dx - \int_{0}^{t} f_{\hat{\tau}^2}(x\mid \hat{\tau}_M^2) ~dx = 0 \right. \right\} \\
    &= F_{\hat{\tau}^2}^{-1} \left( \left. \frac{1}{2} \right|  \hat{\tau}^2_M \right).
\end{align*}
Hence, the parameter $\tau^{*2}$ can be estimated as
\begin{align*}
    \hat{\tau}_M^2 &= \left\{ \tau^2 \in [0, \infty) \left|  F_{\hat{\tau}^2}(\hat{\tau}^2\mid \tau^2) = \frac{1}{2} \right. \right\}.
\end{align*}
\rightline{$\square$}

\subsubsection{Proof of Proposition \ref{prop2}}\label{ap-prop2}
Using Markov's inequality and a first-order Taylor expansion, for any $\epsilon > 0$, we have
\begin{align*}
    P\left(\sup_{y \in (-\infty, \infty)} \left|  \hat{F}_{\hat{\theta}}(y\mid \hat{\tau}_M^2) - F_{\hat{\theta}}(y\mid \tau^{*2}) \right|  > \epsilon \right)
    &< \frac{1}{\epsilon} E\left[ \sup_{y \in (-\infty, \infty)} \left|  \hat{F}_{\hat{\theta}}(y\mid \hat{\tau}_M^2) - F_{\hat{\theta}}(y\mid \tau^{*2}) \right|  \right] \\
    &= \frac{1}{\epsilon} E\left[ \sup_{y \in (-\infty, \infty)} \left|  \int_0^{\infty} F_{\hat{\theta}\mid \hat{\tau}^2}(y\mid x) \left\{ f_{\hat{\tau}^2}(x\mid \hat{\tau}^{2}) - f_{\hat{\tau}^2}(x\mid \tau^{*2}) \right\} \right|  \right] \\
    &\le \frac{1}{\epsilon} E\left[ \left|  \int_0^{\infty} \left\{ f_{\hat{\tau}^2}(x\mid \hat{\tau}^{2}) - f_{\hat{\tau}^2}(x\mid \tau^{*2}) \right\} \right|  \right] \\
    &= \frac{1}{\epsilon} E\left[ \left|  \int_0^{\infty} \left\{ \left. \frac{\partial}{\partial \tau^2} f_{\hat{\tau}^2}(x\mid \tau^{2}) \right| _{\tau^2=c} \right\} (\hat{\tau}_M^2 - \tau^{*2}) \right|  \right] \\
    &< \frac{M}{\epsilon} E[\mid \hat{\tau}_M^2 - \tau^{*2}\mid ],
\end{align*}
where $c$ is a positive constant such that $|c - \tau^{*2}|  < |\hat{\tau}_M^2 - \tau^{*2}|$, and $M = \left|  \int_0^{\infty} \left\{ \left. \frac{\partial}{\partial \tau^2} f_{\hat{\tau}^2}(x\mid \tau^{2}) \right| _{\tau^2=c} \right\} \right| $. From the proposition 4, $\hat{\tau}_M^2 \to_p \tau^{*2}$. Therefore, Proposition \ref{prop2} holds.

\rightline{$\square$}

\subsection{Proof of Theorem \ref{theorem2}}
\label{app-gmm}

From the Theorem of \citet{Box1954}, $Q_a$ is distributed as $\sum_{r=1}^R \lambda_r \chi_{r(1)}^2$, where $\lambda_1, \dots, \lambda_R$ are non-negative eigenvalues of matrix $VW$, $V$ is the variance-covariance matrix of $(\hat{\theta}_1-\hat{\theta}_a, \dots, \hat{\theta}_K-\hat{\theta}_a)$, $W$ is a diagonal matrix with $a_1, \dots, a_K$, and $\chi_{1(1)}^2, \dots, \chi_{R(1)}^2$ are independently and identically distributed as chi-square distributions with 1 degree of freedom.

\begin{align*}
    E[(\hat{\theta}_k-\hat{\theta}_a)^2] &= E\left[ \hat{\theta}_k^2 - 2\hat{\theta}_k \hat{\theta}_a + \hat{\theta}_a^2 \right] \\
    &= \left(\theta^2+(\sigma_k^2+\tau^2) \right) -2 \left( \theta^2 + \frac{a_k(\sigma_k^2+\tau^2)}{A_1^2} \right) + \left( \theta^2 + \frac{1}{A_1^2} \sum_{i=1}^K a_i^2 (\sigma_i^2+\tau^2) \right) \\
    &= (\sigma_k^2+\tau^2) -2 \frac{a_k(\sigma_k^2+\tau^2)}{A_1^2} + \frac{1}{A_1^2} \sum_{i=1}^K a_i^2 (\sigma_i^2+\tau^2) \\
    &= \left(\sigma_k^2 + \frac{1}{A_1^2}\sum_{i=1}^K a_i^2 \sigma_i^2 - \frac{2 a_k \sigma_k^2}{A_1} \right) + \left(1 + \frac{A_2}{A_1^2} - \frac{2a_k}{A_1} \right) \tau^2,
\end{align*}
where $A_r = \sum_{i=1}^K a_i^r$.
We assume $k \ne k'$,
\begin{align*}
    E[(\hat{\theta}_k-\hat{\theta}_a)(\hat{\theta}_{k'}-\hat{\theta}_a)] &= E\left[ \hat{\theta}_k\hat{\theta}_{k'} - \hat{\theta}_k\hat{\theta}_a - \hat{\theta}_{k'}\hat{\theta}_a + \hat{\theta}_a^2 \right] \\
    &= \theta^2 - \left( \theta^2 + \frac{a_k(\sigma_k^2+\tau^2)}{A_1^2} \right) - \left( \theta^2 + \frac{a_{k'}(\sigma_{k'}^2+\tau^2)}{A_1^2} \right) \\
    &+ \left( \theta^2 + \frac{1}{A_1^2} \sum_{i=1}^K a_i^2 (\sigma_i^2+\tau^2) \right) \\
    &= - \frac{a_k(\sigma_k^2+\tau^2)}{A_1^2} - \frac{a_{k'}(\sigma_{k'}^2+\tau^2)}{A_1^2} + \frac{1}{A_1^2} \sum_{i=1}^K a_i^2 (\sigma_i^2+\tau^2) \\
    &= \left( -\frac{a_k \sigma_k^2 + a_{k'}\sigma_{k'}^2}{A_1} + \frac{1}{A_1^2}\sum_{i=1}^K a_i^2 \sigma_i^2 \right) + \left(\frac{A_2}{A_1^2} - \frac{a_k+a_{k'}}{A_1} \right)\tau^2.
\end{align*}
Thus, we can calculate the $(k,k')$ element of $V$ as in \eqref{eq:vij}.

The distribution function of the generalized moment estimator of the between-study variance can be given as
\begin{align*}
    F_{\hat{\tau}_{G,u}^2}(x \mid \tau^2) &= \Pr\left\{ \hat{\tau}^2 \leq x \right\} \\
    &= \Pr\left\{ \frac{Q_a - \left( \sum_{k=1}^K a_k \sigma_k^2 - A_1^{-1} \sum_{k=1}^K a_k^2 \sigma_k^2 \right)}{A_1 - A_2 / A_1} \leq x \right\} \\
    &= \Pr\left\{ Q_a \leq (A_1 - A_2 / A_1) x + \left( \sum_{k=1}^K a_k \sigma_k^2 - A_1^{-1} \sum_{k=1}^K a_k^2 \sigma_k^2 \right) \right\} \\
    &= F_{Q_a} \left( (A_1 - A_2 / A_1) x + \left( \sum_{k=1}^K a_k \sigma_k^2 - A_1^{-1} \sum_{k=1}^K a_k^2 \sigma_k^2 \right) \right).
\end{align*}

Therefore, the density function of $\hat{\tau}_{G,u}^2$ can be given as
\begin{align*}
    f_{\hat{\tau}_{G,u}^2}(x) &= (A_1 - A_2 / A_1) f_{Q_a} \left( (A_1 - A_2 / A_1) x + \left( \sum_{k=1}^K a_k \sigma_k^2 - A_1^{-1} \sum_{k=1}^K a_k^2 \sigma_k^2 \right) \right),
\end{align*}

and $\hat{\tau}_{G,u}^2$ is distributed as
\begin{align*}
    \hat{\tau}_{G,u} \sim \frac{\sum_{r=1}^R \lambda_r \chi_{r(1)}^2 - \left( \sum_{k=1}^K a_k \sigma_k^2 - A_1^{-1} \sum_{k=1}^K a_k^2 \sigma_k^2 \right)}{A_1 - A_2 / A_1}.
\end{align*}

\rightline{$\square$}

\subsection{Proof of Corollary \ref{theorem3}}
\label{ap-sj}
We show that the SJ estimator $\hat{\tau}_{SJ}^2$ is distributed as $\frac{\tau^2}{K-1} \chi_{K-1}^2$.
From the Theorem of \citet{Box1954}, $\hat{\tau}_{SJ}^2$ is distributed as $\sum_{r=1}^R \lambda_r \chi_{r(1)}^2$, where $\lambda_1, \dots, \lambda_R$ are non-negative eigenvalues of matrix $VW$, $V$ is the variance-covariance matrix of $(\hat{\theta}_1-\hat{\theta}_v, \dots, \hat{\theta}_K-\hat{\theta}_v)$, $W$ is a diagonal matrix with $(K-1)^{-1} v_1^{-1}, \dots, (K-1)^{-1} v_K^{-1}$, and $\chi_{1(1)}^2, \dots, \chi_{R(1)}^2$ are independently and identically distributed as chi-square distributions with 1 degree of freedom.

We can calculate as follows, similar to the flow of proof for Theorem 2:
\begin{align*}
    E[(\hat{\theta}_k-\hat{\theta}_v)^2] &= \sigma_k^2+\tau^2 - \frac{1}{W_1(\tau^2)} \\
    E[(\hat{\theta}_k-\hat{\theta}_v)(\hat{\theta}_{k'}-\hat{\theta}_v)] &= - \frac{1}{W_1(\tau^2)}.
\end{align*}
Thus,
\begin{align*}
    \det (\lambda - VW) &= \det \left( \lambda I_K - \frac{\tau^2}{K-1} \left(I_K - \frac{1}{W_1(\tau^2)} \left( \begin{array}{cccc}
        w_1(\tau^2) & w_2(\tau^2) & \cdots & w_K(\tau^2)  \\
        w_1(\tau^2) & w_2(\tau^2) & \cdots & w_K(\tau^2)  \\
        \vdots & \vdots & \vdots & \vdots \\
        w_1(\tau^2) & w_2(\tau^2) & \cdots & w_K(\tau^2)
    \end{array} \right) \right) \right) \\
    &= \left|  \begin{array}{cccc}
        \lambda-\frac{\tau^2}{K-1}\left(1-\frac{w_1(\tau^2)}{W_1(\tau^2)}\right) & -\frac{w_2(\tau^2)}{W_1(\tau^2)} & \cdots & -\frac{w_K(\tau^2)}{W_1(\tau^2)}  \\
        -\frac{w_1(\tau^2)}{W_1(\tau^2)} & \lambda-\frac{\tau^2}{K-1}\left(1-\frac{w_2(\tau^2)}{W_1(\tau^2)}\right) & \cdots & -\frac{w_K(\tau^2)}{W_1(\tau^2)} \\
        \vdots & \vdots & \vdots & \vdots \\
        -\frac{w_1(\tau^2)}{W_1(\tau^2)} & -\frac{w_2(\tau^2)}{W_1(\tau^2)} & \cdots & \lambda-\frac{\tau^2}{K-1}\left(1-\frac{w_K(\tau^2)}{W_1(\tau^2)}\right)
    \end{array} \right|  \\
    &= \lambda \left|  \begin{array}{cccc}
        1 & -\frac{w_2(\tau^2)}{W_1(\tau^2)} & \cdots & -\frac{w_K(\tau^2)}{W_1(\tau^2)}  \\
        1 & \lambda-\frac{\tau^2}{K-1} & \cdots & 0 \\
        \vdots & \vdots & \vdots & \vdots \\
        1 & -\frac{w_2(\tau^2)}{W_1(\tau^2)} & \cdots & \lambda-\frac{\tau^2}{K-1}
    \end{array} \right|  \\
        &= \lambda \left|  \begin{array}{cccc}
        1 & -\frac{w_2(\tau^2)}{W_1(\tau^2)} & \cdots & -\frac{w_K(\tau^2)}{W_1(\tau^2)}  \\
        0 & \lambda-\frac{\tau^2}{K-1} & \cdots & 0 \\
        \vdots & \vdots & \vdots & \vdots \\
        0 & 0 & \cdots & \lambda-\frac{\tau^2}{K-1}
    \end{array} \right|  \\
    &= \lambda \left(\lambda - \frac{\tau^2}{K-1} \right)^{K-1}.
\end{align*}
Therefore, we can get $\lambda_r=\frac{\tau^2}{K-1}$ $(r=1, \dots, K-1)$, and the distribution of $\hat{\tau}_{SJ}^2$ can be written as
\begin{align*}
    \hat{\tau}_{SJ}^2 &\sim \sum_{r=1}^R \lambda_r \chi_{r(1)}^2 = \sum_{r=1}^R \frac{\tau^2}{K-1} \chi_{r(1)}^2 = \frac{\tau^2}{K-1} \chi_{K-1}^2.
\end{align*}

\rightline{$\square$}

\section{Supplementary simulations}

\subsection{Simulation result of HKSJ approach}
\label{ap-hksj}
\begin{figure}[ht]
    \centerline{
    \includegraphics[width=180mm]{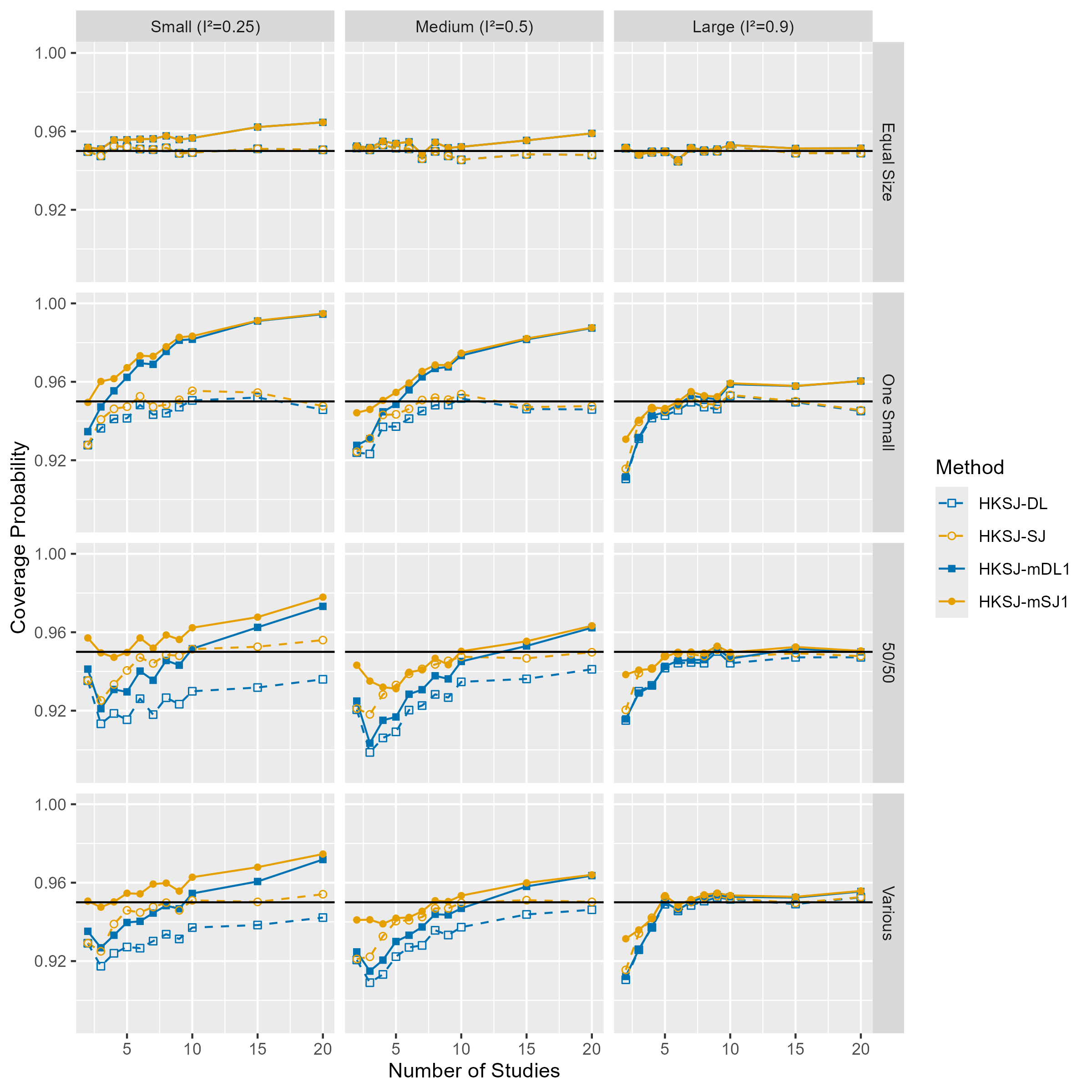}    
    }
    \caption{Coverage probabilities of the overall treatment effect by the HKSJ approach with the DL estimator (HKSJ-DL), the SJ estimator (HKSJ-SJ), and proposed M-estimator method (HKSJ-mDL1, HKSJ-mSJ1). Rows represent within-study variance settings, and columns represent between-study heterogeneity $I^2=0.25, 0.5, 0.9$.}
    \label{fig:cp-theta-HKSJ}
\end{figure}

\begin{figure}[ht]
    \centerline{
    \includegraphics[width=180mm]{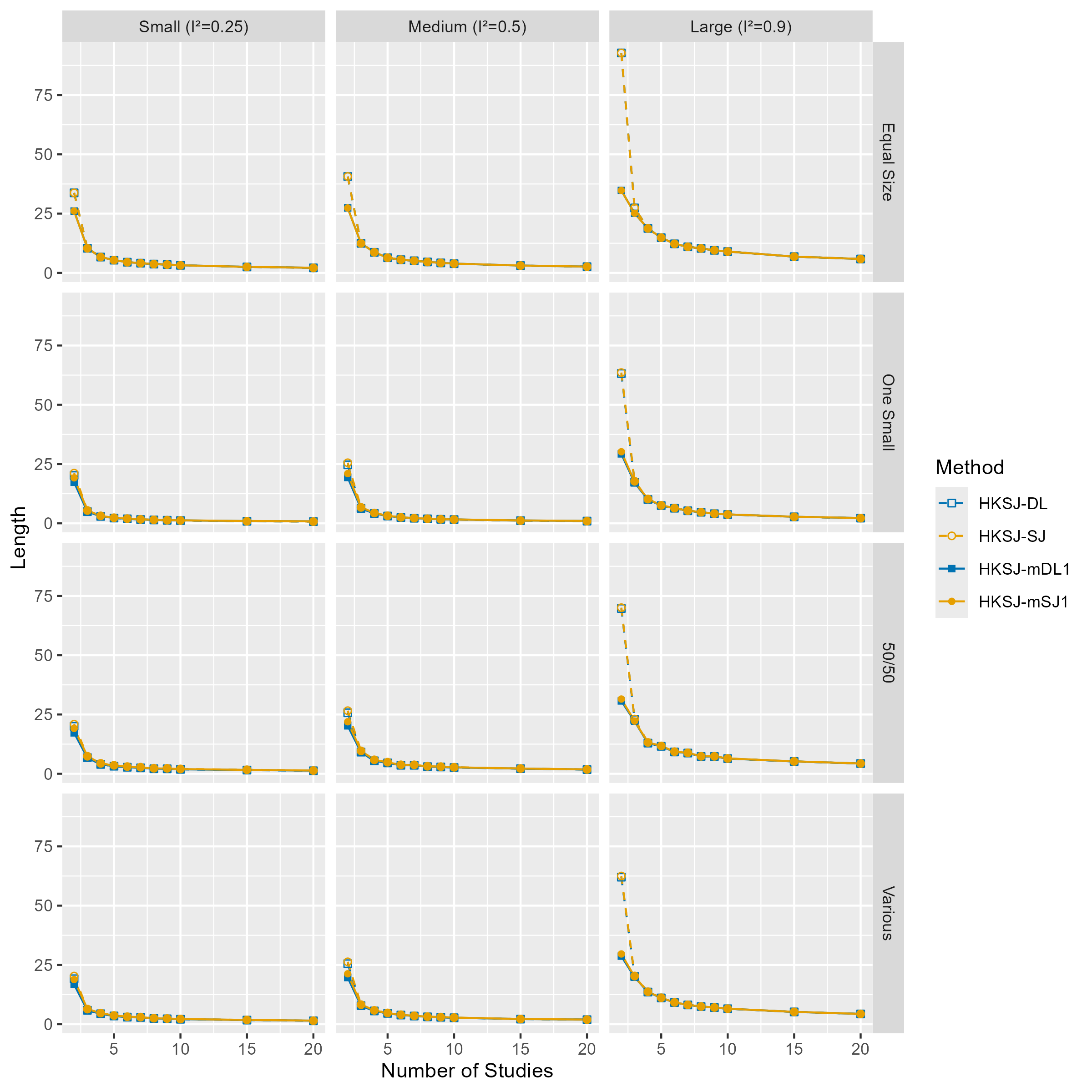}    
    }
    \caption{Lengths of the overall treatment effect by the HKSJ approach with the DL estimator (HKSJ-DL), the SJ estimator (HKSJ-SJ), and proposed M-estimator method (HKSJ-mDL1, HKSJ-mSJ1). Rows represent within-study variance settings, and columns represent between-study heterogeneity $I^2=0.25, 0.5, 0.9$.}
    \label{fig:length-theta-HKSJ}
\end{figure}

\begin{figure}[ht]
    \centerline{
    \includegraphics[width=180mm]{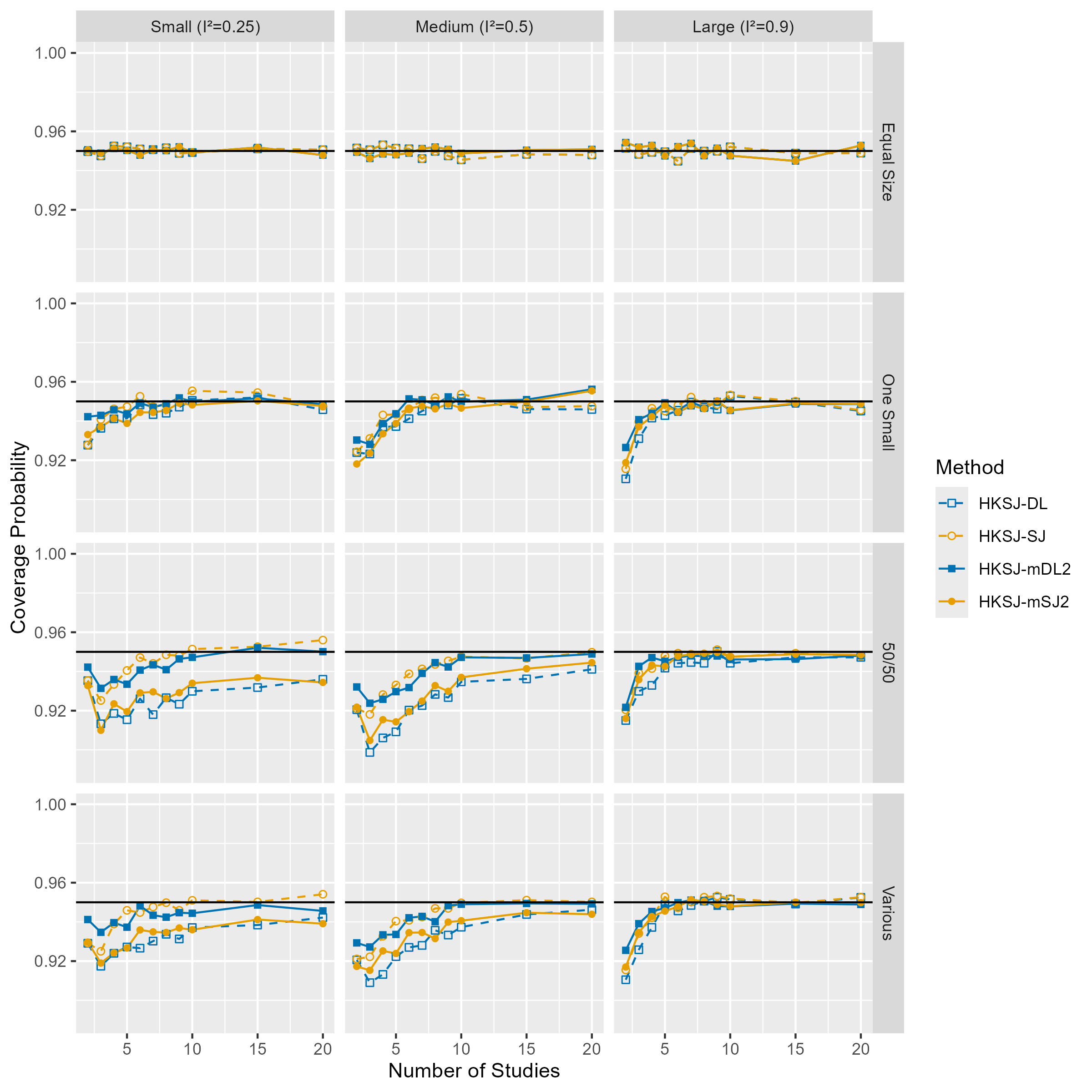}    
    }
    \caption{Coverage probabilities of the overall treatment effect by the HKSJ approach with the DL estimator (HKSJ-DL), the SJ estimator (HKSJ-SJ), and proposed M-estimator method (HKSJ-mDL2, HKSJ-mSJ2). Rows represent within-study variance settings, and columns represent between-study heterogeneity $I^2=0.25, 0.5, 0.9$.}
    \label{fig:cp-theta-HKSJ-M_2}
\end{figure}

\begin{figure}[ht]
    \centerline{
    \includegraphics[width=180mm]{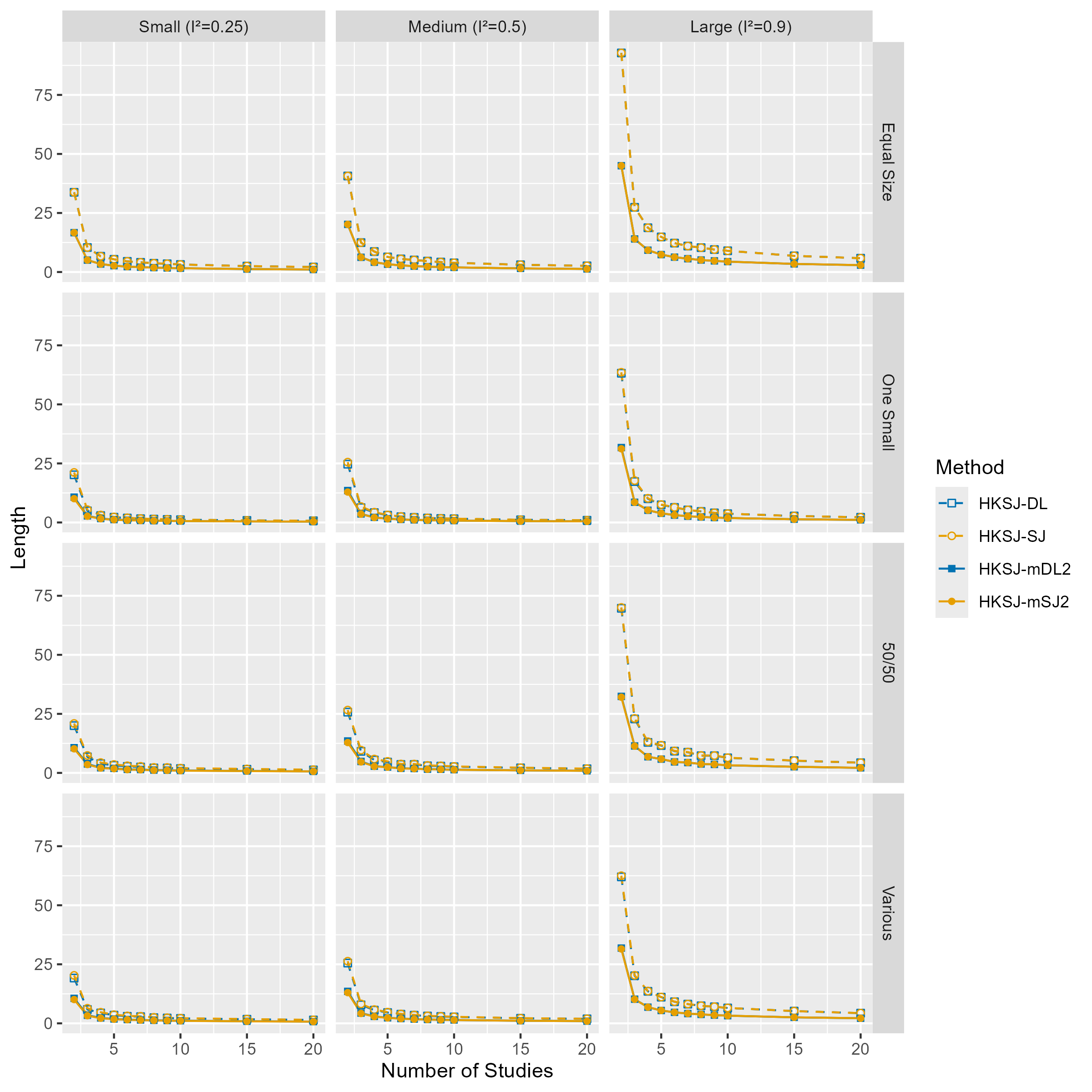}
    }
    \caption{Lengths of the overall treatment effect by the HKSJ approach with the DL estimator (HKSJ-DL), the SJ estimator (HKSJ-SJ), and proposed M-estimator method (HKSJ-mDL2, HKSJ-mSJ2). Rows represent within-study variance settings, and columns represent between-study heterogeneity $I^2=0.25, 0.5, 0.9$.}
    \label{fig:length-theta-HKSJ-M_2}
\end{figure}

The coverage probabilities and lengths of the CI for $\theta$ by the HKSJ approach with comparison methods (HKSJ-DL, HKSJ-SJ) and our proposed one (HKSJ-mDL1, HKSJ-mDL2, HKSJ-mSJ1, HKSJ-mSJ2) are shown in Figures \ref{fig:cp-theta-HKSJ}-\ref{fig:length-theta-HKSJ-M_2}.
As with typical methods, our proposed approach consistently demonstrated a higher coverage probability than the HKSJ approach with the DL and SJ estimators, particularly showing a significant improvement for the DL estimator. The HKSJ-mDL2 approach consistently approximated the nominal level more closely than the existing HKSJ approach. While HKSJ-mDL1 and HKSJ-mSJ1 reached the nominal level with a small number of studies and low between-study heterogeneity, these methods tended to be overly conservative as the number of studies increased and heterogeneity remained low.
Furthermore, the correction applied with the M-estimator approach did not result in excessive conservative, regardless of the number of studies or the level of between-study heterogeneity. Overall, these results suggest that HKSJ-mDL2 is advantageous across a range of settings.

\subsection{The number of computational failures for estimating CI}\label{ap-nfail}
\begin{figure}[ht]
    \centerline{
    \includegraphics[width=180mm]{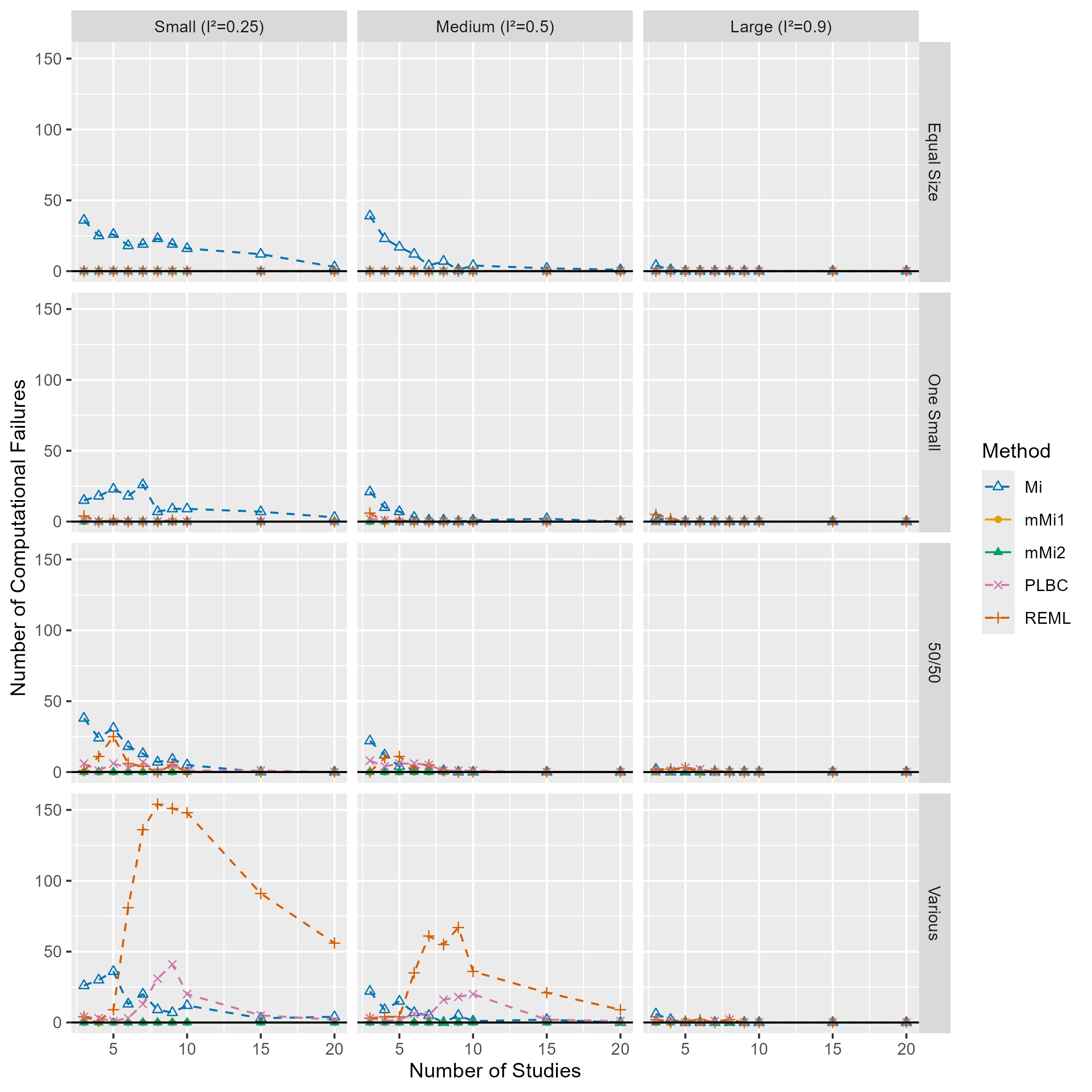}    
    }
    \caption{Number of computational failures regarding the CI of the overall treatment effect by \citet{michael_exact_2019} (Mi), the proposed M-estimator method (mMi1, mMi2), the restricted maximum likelihood method (REML), and the profile likelihood method with Bartlett-type corrections (PLBC). Rows represent within-study variance settings, and columns represent between-study heterogeneity $I^2 = 0.25, 0.5, 0.9$. The number of iterations is 10,000.}
    \label{fig:N-michael}
\end{figure}

The numbers of computational failures related to the CI for the overall treatment effect by the Mi, mMi, REML, and PLBC methods are illustrated in Figure \ref{fig:N-michael}. The Mi method is feasible for estimating the CI of the overall treatment effect when the number of studies is sufficiently large, but it can become infeasible in rare cases with a small number of studies. This issue arises when the CI for the between-study variance cannot be computed. Since the CI for the overall treatment effect depends on the search range for the between-study variance, if the CI for the between-study variance fails to compute, the CI for the overall treatment effect cannot be calculated either. In such cases, the mMi method offers an advantage, as it can compute the CI without computational failures.

Both the REML and PLBC methods also encountered computational failures under various within-study variance settings. These failures occur because, in certain scenarios, no solution exists for these methods.

\bibliography{bibitems}
\bibliographystyle{elsarticle-harv}




\end{document}